\documentclass[aps,prl,twocolumn,showpacs,preprintnumbers,amsmath,amssymb]{revtex4-1}

\usepackage{braket}
\usepackage{amsmath}
\usepackage{amsthm}
\usepackage{amssymb}

 \def\frac#1#2{{#1\over #2}}

 \def\s{\sqrt}

\def\be{\begin{equation}}
\def\ee{\end{equation}}
\def\ba{\begin{eqnarray}}
\def\ea{\end{eqnarray}}
\newcommand{\Tr}{\text{Tr}}

 \def\de{\partial}

 \def\ti{\tilde}

 \def\ddd{\cdot\cdot\cdot}
 \def\no{\nonumber \\}

 \def\la{\langle}
 \def\lb{\rangle}

\usepackage{color}
\usepackage{graphicx}
\usepackage{dcolumn}
\usepackage{bm}
\usepackage{hyperref}

\begin{document}

\title{Entanglement of Purification in Many Body Systems and Symmetry Breaking}
\begin{flushleft}
YITP-19-05 ; IPMU19-0014
\end{flushleft}
\author{Arpan Bhattacharyya$^{a}$,
Alexander Jahn$^{b}$, Tadashi Takayanagi$^{a,c}$ and Koji Umemoto$^{a}$}

\affiliation{$^a$Center for Gravitational Physics,
Yukawa Institute for Theoretical Physics,\\
Kyoto University, 
Kitashirakawa Oiwakecho, Sakyo-ku, Kyoto 606-8502, Japan}

\affiliation{$^b$Dahlem Center for Complex Quantum Systems,\\
 Freie Universit{\"a}t Berlin,
 14195 Berlin, Germany}

\affiliation{$^{c}$Kavli Institute for the Physics and Mathematics
 of the Universe (WPI),\\
University of Tokyo, Kashiwa, Chiba 277-8582, Japan}

\date{\today}

\begin{abstract}

We study the entanglement of purification (EoP), a measure of total correlation between two subsystems $A$ and $B$, for free scalar field theory on a lattice and the transverse-field Ising model by numerical methods. 
In both of these models, we find that the EoP becomes a non-monotonic function of the distance between $A$ and $B$ when the total number of lattice sites is small. When it is large, the EoP becomes monotonic and shows a plateau-like behavior. Moreover, we show that the original reflection symmetry which exchanges $A$ and $B$ can get broken in optimally purified systems. In the Ising model, we find this symmetry breaking in the ferromagnetic phase.
We provide an interpretation of our results in terms of the interplay between classical and quantum correlations.
\end{abstract}

\maketitle

{\bf 1. Introduction}

The entanglement entropy (EE) is  a unique measure of quantum entanglement for pure states \cite{EEunique}. Decomposing a total quantum system into two subsystems $A$ and $B$, the EE is defined as
$S_A=-\mbox{Tr}[\rho_A\log\rho_A]$, where the reduced density matrix is
$\rho_A\equiv \mbox{Tr}_B|\Psi \lb_{AB}\la \Psi|_{AB}$, and $|\Psi\lb_{AB}$ describes a pure state. The EE helps us to extract essential properties of quantum field theories
\cite{BKLS,Sr}, especially conformal field theories (CFTs) \cite{HLW}.
It has recently played an important role in the context of the holographic Anti de-Sitter space/conformal field theory (AdS/CFT) correspondence \cite{Ma}, due its simple geometrical interpretation in gravity \cite{RT,HRT}.

Quantities such as entanglement of formation and squashed entanglement extend EE to mixed states, where the EE itself is not a good measure of quantum entanglement or classical correlations (refer to e.g. a comprehensive review \cite{HHHH}). However, such quantities often require a minimization over infinitely many quantum states and are thus computationally challenging in quantum field theory, leading to a scarcity of results.

This letter provides a first step toward such a minimization. We will study entanglement of purification (EoP) $E_P(\rho_{AB})$ \cite{EP,BP}, a simpler version of more complicated mixed state entanglement measures and defined as follows: Consider a purification $|\Psi\lb_{A\ti{A}B\ti{B}}$ of a mixed state $\rho_{AB}$, i.e. a pure state in an enlarged Hilbert space ${\cal H}_A\otimes {\cal H}_B \to {\cal H}_A\otimes {\cal H}_B \otimes {\cal H}_{\ti{A}} \otimes {\cal H}_{\ti{B}}$ with a constraint
\be
\mbox{Tr}_{\ti{A}\ti{B}}\left[|\Psi\lb_{A\ti{A}B\ti{B}}\la\Psi|_{A\ti{A}B\ti{B}}\right] = \rho_{AB}\ .
\label{condw}
\ee
EoP is given by the minimal EE $S_{A\ti{A}}$ over all purifications $|\Psi\lb_{A\ti{A}B\ti{B}}$:
\ba
E_P(\rho_{AB})=\min_{|\Psi\lb_{A\ti{A}B\ti{B}}}S_{A\ti{A}}\ .  \label{EoPdef}
\ea

EoP is a measure of total correlation between the two subsystems $A$ and $B$: It vanishes only for product states and monotonically decreases under local operations, while its regularization possesses an operational meaning in terms of EPR pairs \cite{EP}. 
Moreover, an AdS/CFT-based geometric interpretation was conjectured \cite{UT,Nguyen:2017yqw}, supported by CFT approaches for specific examples \cite{HEoPCFT}, and actively studied \cite{HEP1,HEP2,HEP3,HEP4,HEP5,HEP6,HEP7,HEP8,HEP9,HEP10,HEP11,
HEP12,HEP13,HEP14,HEP15,HEP16,HEP17,HEP18,HEP19,HEP20,HEP21,HEP22}, motivating a field-theoretic treatment. Earlier work on EoP for free scalar field theory has been performed for small subsystems \cite{Bhattacharyya:2018sbw}.

In this letter, we numerically study the EoP in free scalar field theory for larger subsystems assuming a Gaussian ansatz, as well as in the transverse-field Ising chain.
Both models exhibit intriguing non-monotonic and plateau-like behavior of EoP with respect to the distance between the subsystems. 
Moreover, we observe a breaking of the $Z_2$ reflection symmetry that exchanges $A\ti{A}$ and $B\ti{B}$ for an optimal purification, 
reminiscent of spontaneous symmetry breaking and unobserved in previous work \cite{Bhattacharyya:2018sbw}.\\

{\bf 2. EoP in free scalar field theory}

Consider a lattice free scalar field theory in $1+1$ dimensions, defined by the Hamiltonian
\be
H=\frac{1}{2}\int^\infty_{-\infty} dx \left[\pi^2+(\de_x\phi)^2+m^2\phi^2\right]\ .
\ee
The ground state wave function $\Psi_0$ for this theory is Gaussian
 \cite{BKLS,Sha,Bhattacharyya:2018sbw},
\be
\Psi_0[\phi]={\cal N}_0\cdot e^{-\frac{1}{2}\sum^N_{n,n'=1}\phi_n' W_{n n'}\phi_{n'}'}
\equiv {\cal N}_0\cdot e^{-\frac{1}{2}\phi^T W\phi}\ . \label{EQ_PHI0}
\ee
The matrix $W$ is defined by
\be
W_{nn'}=\frac{1}{N}\sum_{k=1}^N\s{m^2a^2+4\sin^2\left(\frac{\pi k}{N}\right)}e^{\frac{2\pi ik(n-n')}{N}}\ ,
\label{EQ_W}
\ee
where $N$ is the total number of lattice sites. 
We set the lattice spacing $a=1.$
Notice that $W$ is symmetric and real-valued. We consider masses between $m=10^{-1}$ and $m=10^{-4}$ near the conformal (massless) limit.

\begin{figure}
  \centering
  \includegraphics[width=3cm]{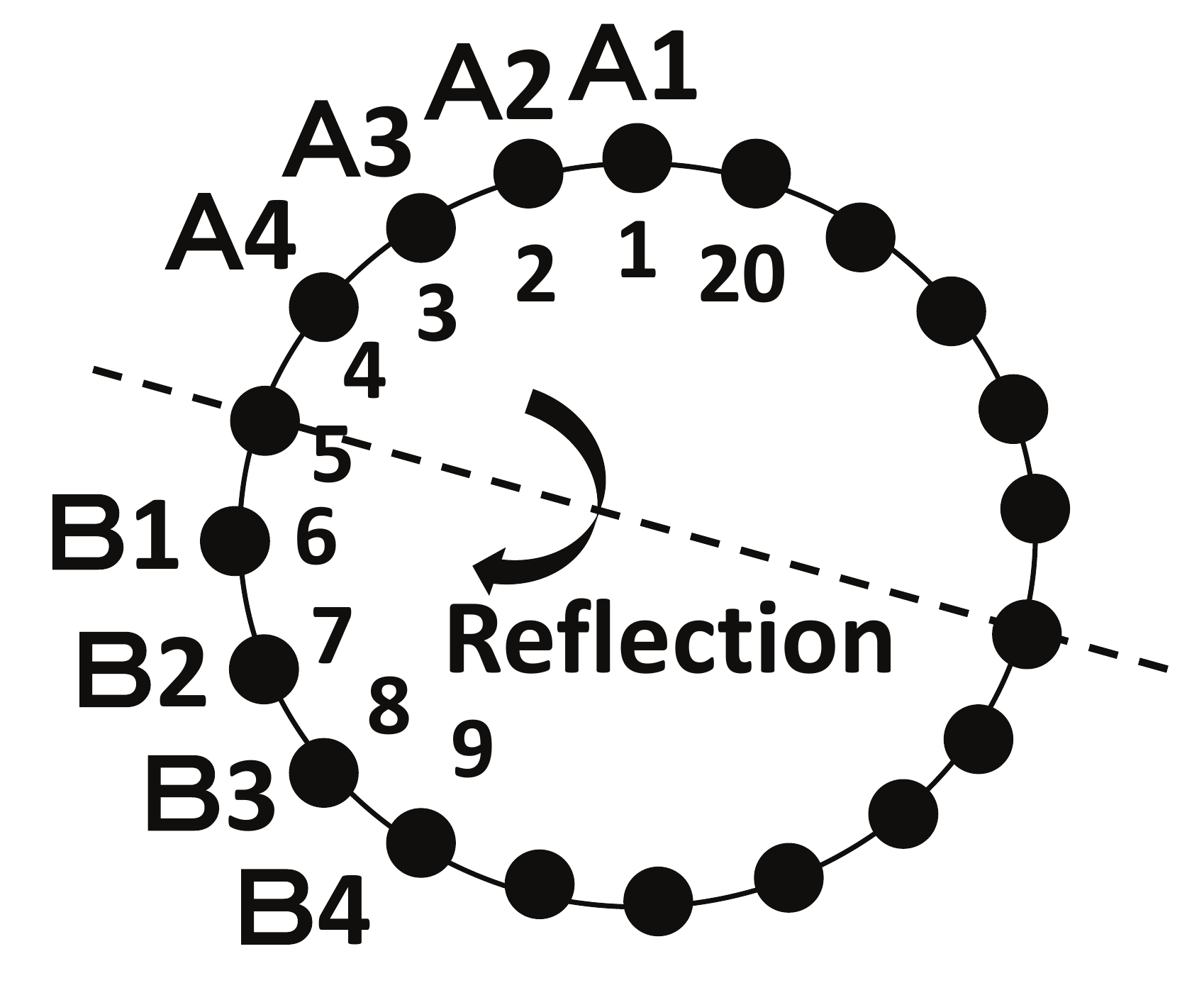}
  \caption{An example of the setup for our lattice model with $N=20$ and $|A|=|B|=4$.
  The distance between $A$ and $B$ is $d=1$. There is an $Z_2$ reflection symmetry.}
\label{fig:setup}
  \end{figure}
  
We divide the total Hilbert space into three parts $\mathcal{H}_\text{tot} = \mathcal{H}_A \otimes \mathcal{H}_B \otimes \mathcal{H}_C$ (Fig.\ \ref{fig:setup}). We denote the number of lattice sites in $A,B$ by $|A|,|B|$ and the distance between them by $d$.
Then \eqref{EQ_PHI0} is written as
\begin{equation}
\label{EQ_PHI0_B}
\Psi_0[\phi] = \mathcal{N}_0\, \exp \left[ -\frac{1}{2}
\begin{pmatrix}
\phi_{AB} \\
\phi_C
\end{pmatrix}^\mathrm{T}
\begin{pmatrix}
P & Q \\
Q^\mathrm{T} & R
\end{pmatrix}
\begin{pmatrix}
\phi_{AB} \\
\phi_C
\end{pmatrix}
\right]
 \text{ ,}
\end{equation}
with the sub-matrices $P,Q,R$ determined by \eqref{EQ_W}.

\begin{figure}[tb]
  \centering
  \includegraphics[width=3.6cm]{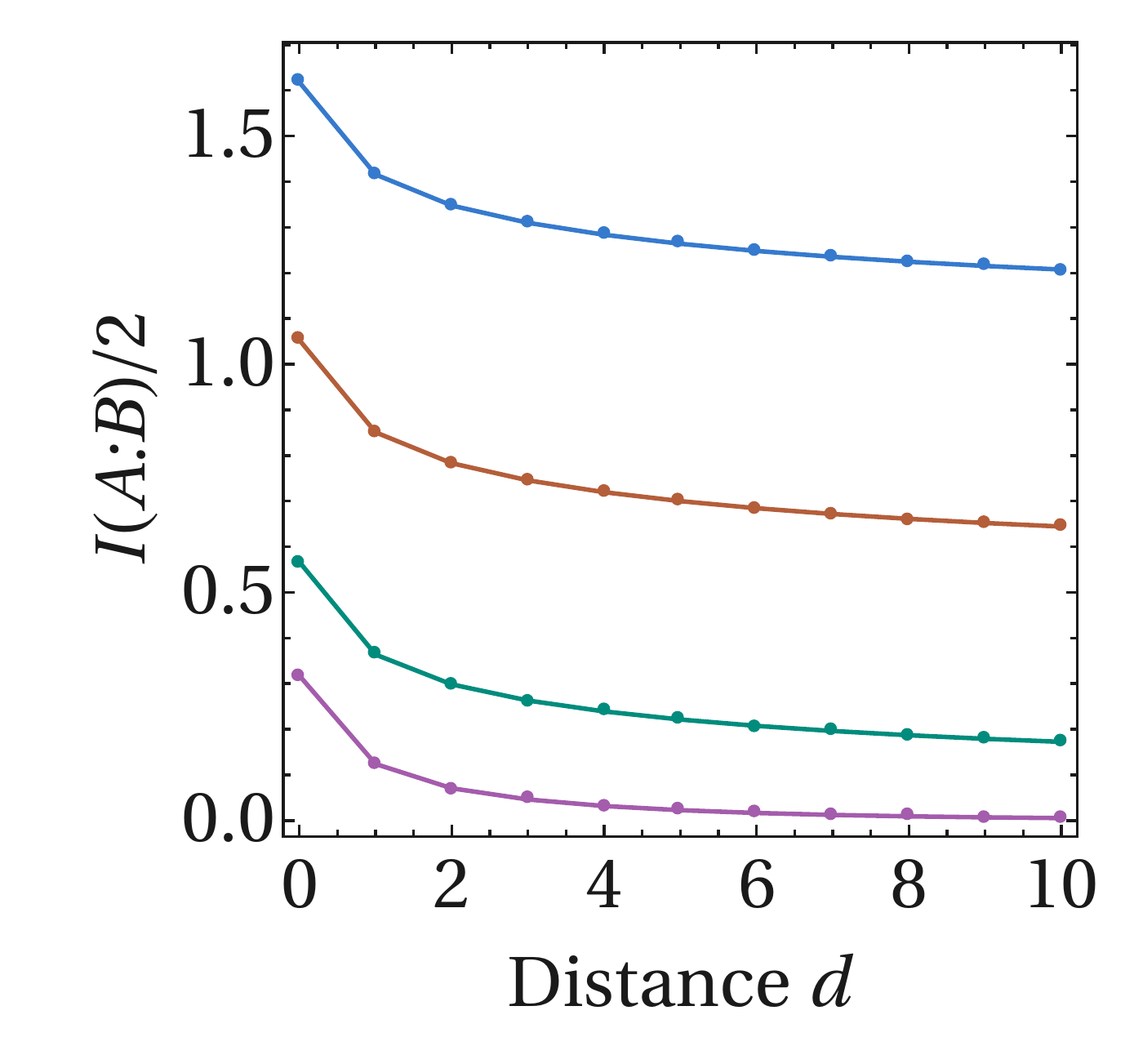}
  \hspace{0.2cm}
  \includegraphics[width=3.6cm]{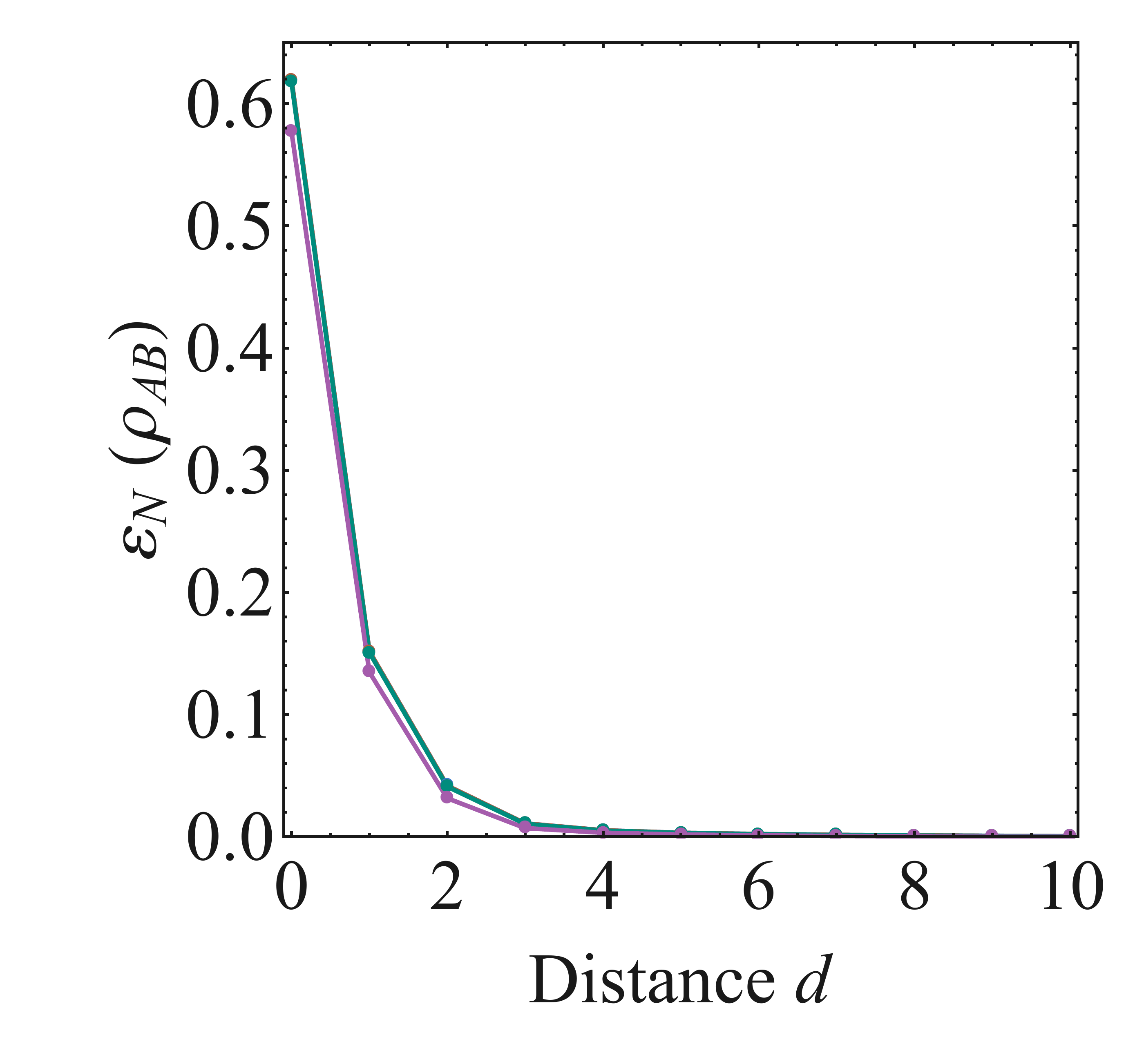}
  \caption{Half of MI (left) and LN (right) for $|A|=|B|=4$ and $N=60$ as a function of $d$, shown for mass $m=10^{-1},10^{-2},10^{-3},10^{-4}$ (bottom to top).}
\label{fig:NGB}
\end{figure}

From this wave functional, we can compute the mutual information (MI) $I(A:B)=S_A+S_B-S_{AB}$ 
and the logarithmic negativity (LN) $\mathcal{E}_{N}(\rho_{AB})$, both of which are shown in Fig.\ \ref{fig:NGB}.
MI is a measure of total correlation satisfying $I(A:B)/2\leq E_P(\rho_{AB})$ \cite{BP}. LN is a useful probe of quantum entanglement between $A$ and $B$ \cite{JensEisertPhD,Audenaert2002}, defined 
as $\mathcal{E}_{N}(\rho_{AB})=\log \Tr |\rho_{AB}^{\Gamma_{B}}|$ \cite{VW,Plenio}, 
where $\rho_{AB}^{\Gamma_{B}}$ is the partial transposition with respect to $B$. Refer to Appendix A for the details of computing $\mathcal{E}_{N}(\rho_{AB})$.
We observe that $\mathcal{E}_N(\rho_{AB})$ takes the largest value at $d=0$ and for $d\geq 1$ shows exponential decay. 
On the other hand, MI slowly decreases as function of $d$ (refer to Appendix B for the scaling law of MI and EoP in the conformal limit).

To calculate the EoP, we purify the system by adding auxiliary subsystems $\tilde{A}$ and $\tilde{B}$. 
Assuming the purified wave functional is Gaussian, we obtain
\begin{align}
\label{EQ_PHIAB}
&\Psi_{A\tilde{A}B\tilde{B}}[\phi] = \mathcal{N}_{A\tilde{A}B\tilde{B}}\, \exp \left( -\frac{1}{2} \phi^\mathrm{T} V \phi \right) \nonumber \\
&\;= \mathcal{N}_{A\tilde{A}B\tilde{B}}\, \exp \left[ -\frac{1}{2}
\begin{pmatrix}
\phi_{AB} \\
\phi_{\tilde{A}\tilde{B}}
\end{pmatrix}^\mathrm{T}
\begin{pmatrix}
J & K \\
K^\mathrm{T} & L
\end{pmatrix}
\begin{pmatrix}
\phi_{AB} \\
\phi_{\tilde{A}\tilde{B}}
\end{pmatrix}
\right]
\text{ ,}
\end{align}
where we have decomposed the matrix $V$ into three sub-matrices $J,K,L$. 
The condition (\ref{condw}) requires $J=P$. Furthermore, assuming subsystems of equal width $w=|A|=|B|$, and setting $|\ti{A}|=|\ti{B}|=w$, $L$ becomes a $2w \times 2w$ square matrix and is related to $K$ by the equation
\begin{equation}
\label{EQ_LMATRIX}
L^{-1} = (K^{-1} Q) R^{-1} (K^{-1} Q)^\mathrm{T} \text{ .}
\end{equation}
Use of a symmetry transformation \cite{Bhattacharyya:2018sbw} allows the simplification of the $K$ to the form:
\begin{equation}
\label{EQ_KMATRIX}
K=
\begin{pmatrix}
1_{w} & K_{A, \tilde{B}} \\
K_{B, \tilde{A}} & 1_{w}
\end{pmatrix}
\text{ .}
\end{equation}
Thus, all parameters of the purification are contained in the $w \times w$ matrices $K_{A, \tilde{B}}$ and $K_{B, \tilde{A}}$. If one assumes a $Z_2$ symmetry which reflects $A\ti{A}$ and $B\ti{B}$, we will have $K_{A, \tilde{B}} =K_{B, \tilde{A}}^\mathrm{R}$, where we define  $M^\mathrm{R}$ of a matrix $M$ as the inverse ordering of all rows and columns, i.e.
\begin{equation}
(K_{B, \tilde{A}}^\mathrm{R})_{j,k} = (K_{B, \tilde{A}})_{w+1-j,w+1-k}\ .
\end{equation}
The $Z_2$ {\it asymmetry} $\mathcal{A}$ is defined to quantify the $Z_2$ symmetry breaking as
\begin{equation}
\label{EQ_Z2_ASYM}
\mathcal{A} = ||K_{A, \tilde{B}} - K_{B, \tilde{A}}^\mathrm{R}||_2 \ ,
\end{equation}
where $||M||_2$ is the 2-norm over all entries of $M$. The actual value of $E_P$ is $Z_2$-invariant due to $S_{A\ti{A}}=S_{B\ti{B}}$.  

Then $S_{A\tilde{A}}$ can be computed from the eigenvalue spectrum $\{\lambda_k \}$
of the matrix $\Lambda \equiv -V^{-1}_{A\tilde{A},B\tilde{B}}\cdot V^{\phantom{-1}}_{B\tilde{B},A\tilde{A}}$ 
\cite{BKLS} as follows:
\begin{equation}
S_{A\tilde{A}} = \sum_k \left( \log\frac{\sqrt{\lambda_k}}{2} + \sqrt{1+\lambda_k} \log\frac{1+\sqrt{1+\lambda_k}}{\lambda_k} \right) \text{ .}
\end{equation}
The EoP is the minimum of $S_{A\tilde{A}}$ over all purifications $\Psi_{A\tilde{A}B\tilde{B}}[\phi]$, achieved by varying $K_{A, \tilde{B}}$ and $K_{B, \tilde{A}}$. 
We computed the EoP for subsystem sizes $w=1,2,3,4$ and studied its dependence on the distance $d$, using a numerical L-BFGS optimization implemented with the C++ package \textsl{dlib} \cite{dlib}.
Here we made the assumption that an optimal purification is contained in the Gaussian purification above. 
We also assumed that the auxiliary subsystems have the same sizes as the original ones, and larger numerical setups did not appear to reduce the optimal EoP further.
If either assumption were inaccurate, our results for free scalar field theory would only provide an upper bound on the EoP.

\begin{figure}[tb]
\centering
\begin{align*}
\hspace{0.5cm}
\begin{gathered}
\includegraphics[height=2.6cm]{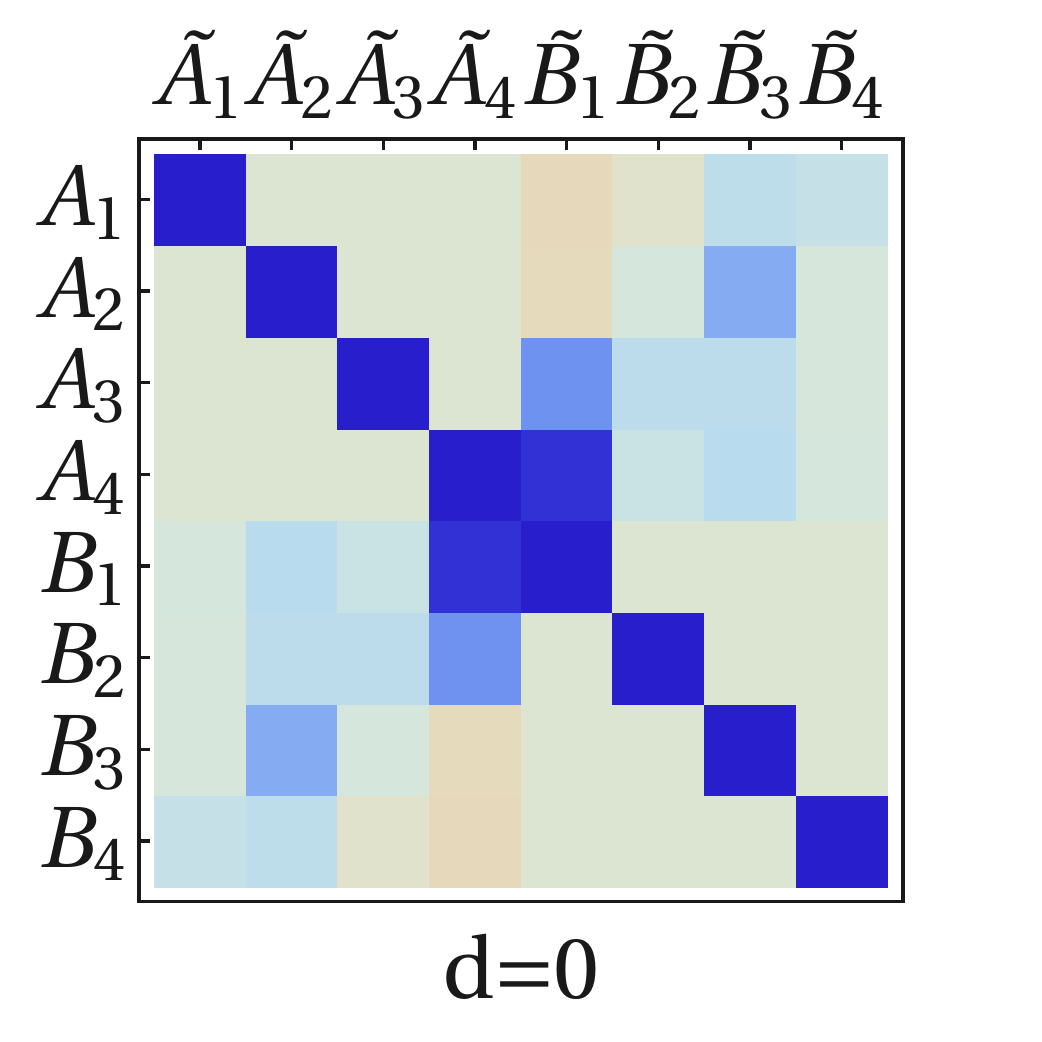}\hspace{0.5cm}
\includegraphics[height=2.6cm]{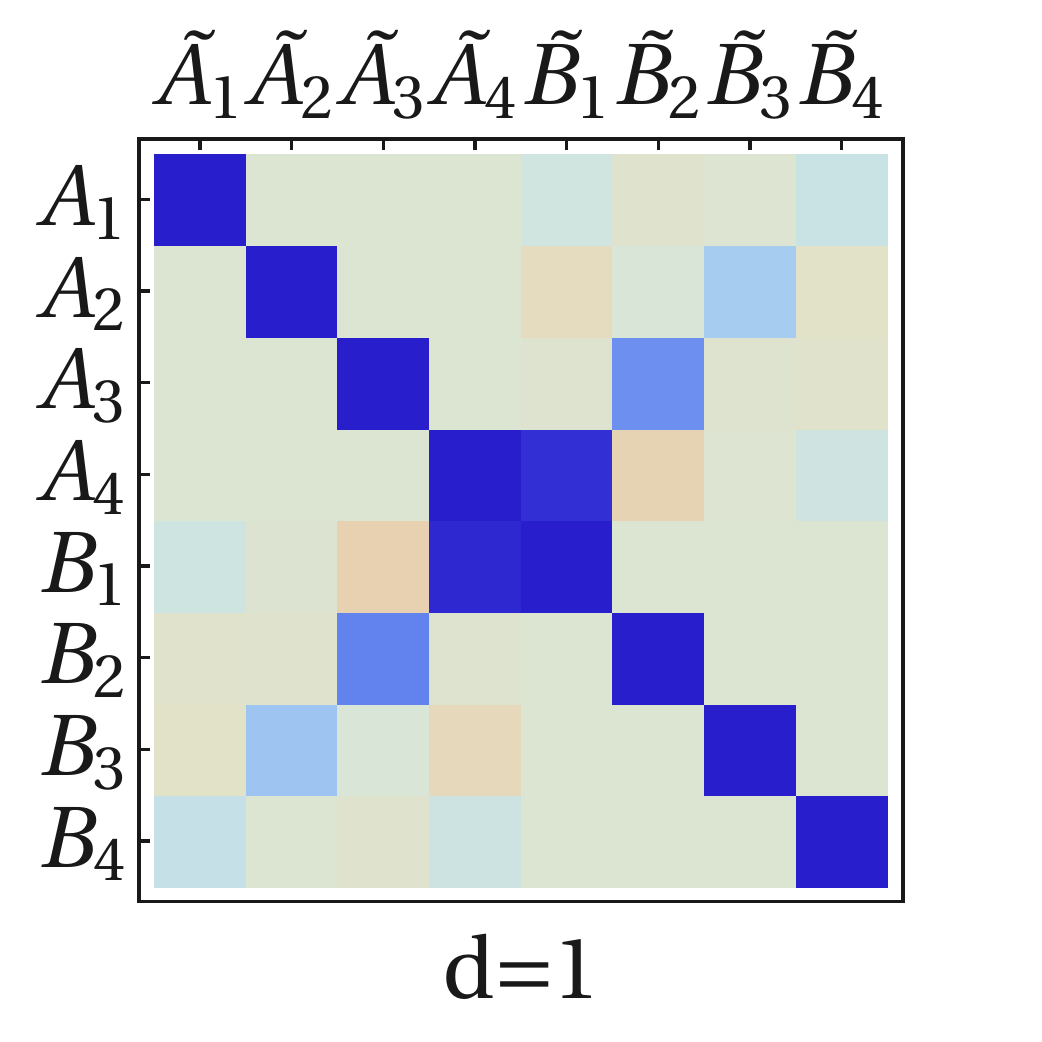}\\
\includegraphics[height=2.6cm]{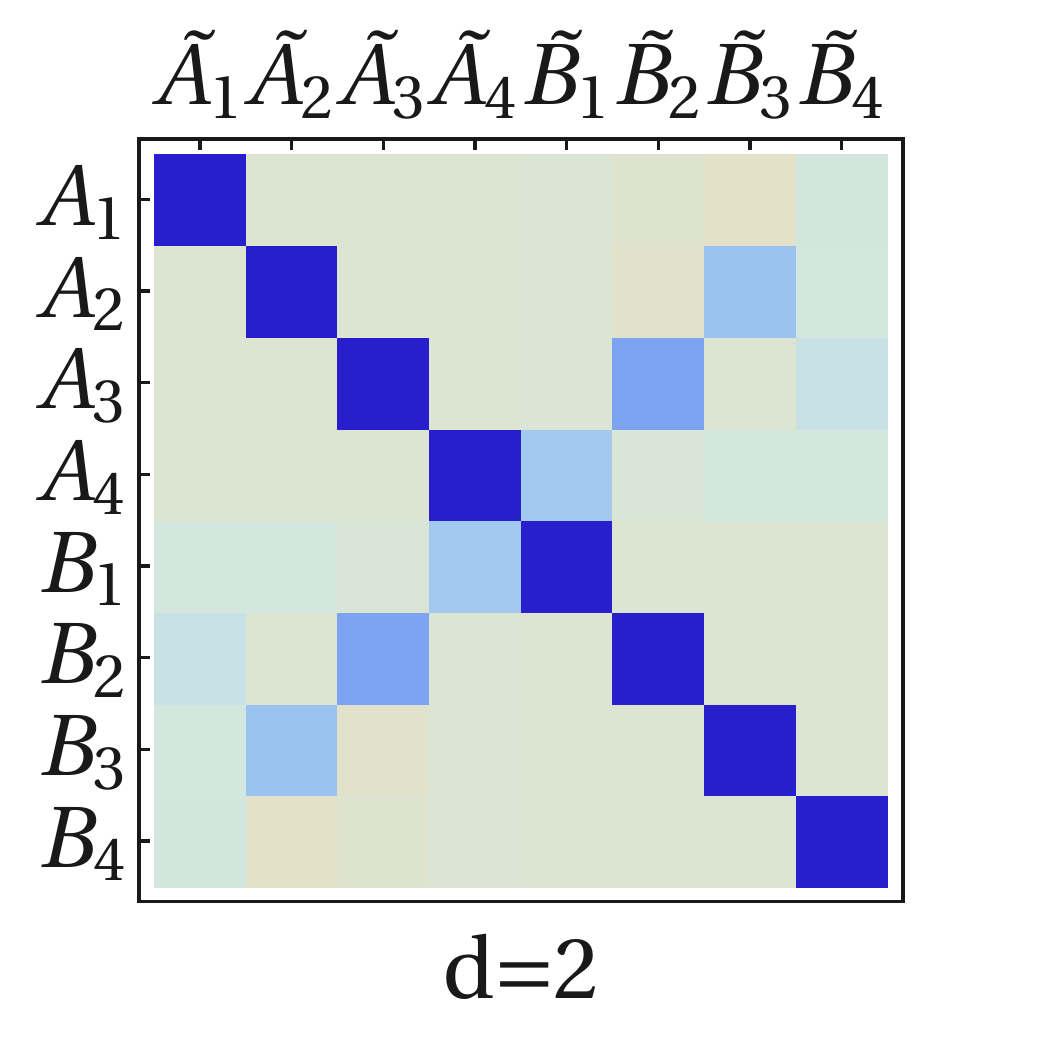}\hspace{0.5cm}
\includegraphics[height=2.6cm]{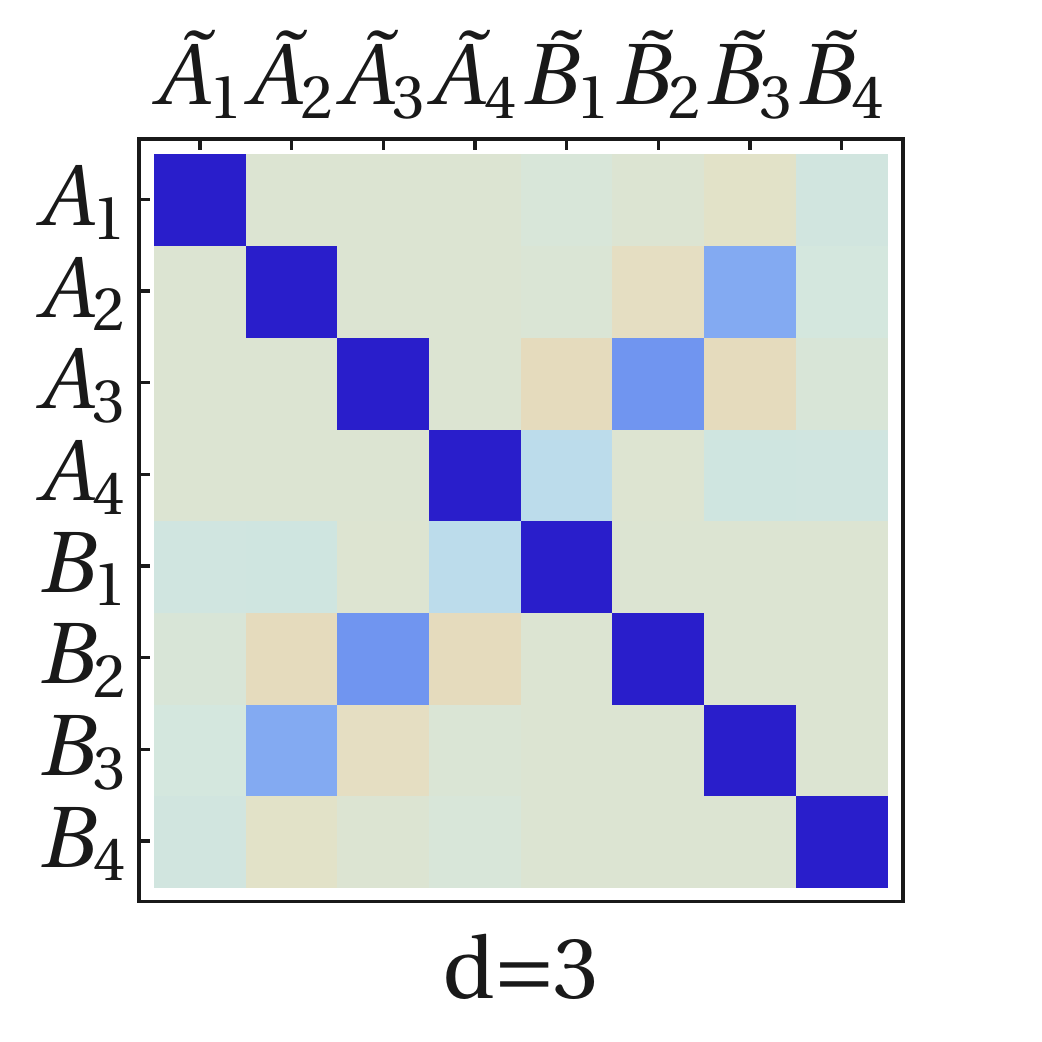}
\end{gathered}
\hspace{0.25cm}
\begin{gathered}
\includegraphics[width=0.55cm]{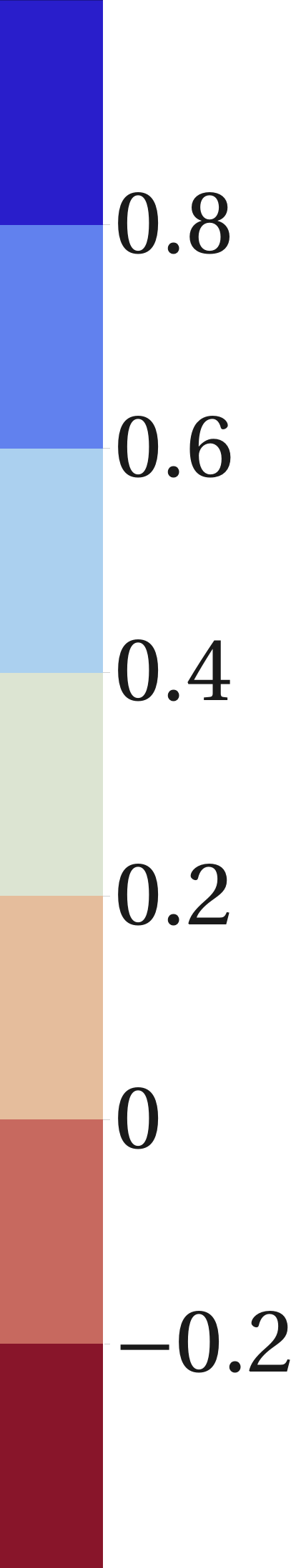}
\end{gathered}
\end{align*}
\vspace{-0.5cm}
\caption{Coupling matrix $K$ (defined in \eqref{EQ_KMATRIX}) for minimal entanglement 
of purification between physical sites $AB$ and auxiliary sites $\tilde{A}\tilde{B}$ for mass $m=10^{-4}$, 
block width $w=|A|=|B|=4$, $N=60$ and various block distances $d$.
}
\label{FIG_KMATRIX}
\end{figure}

\begin{figure}[tb]
\centering
\includegraphics[width=3.4cm]{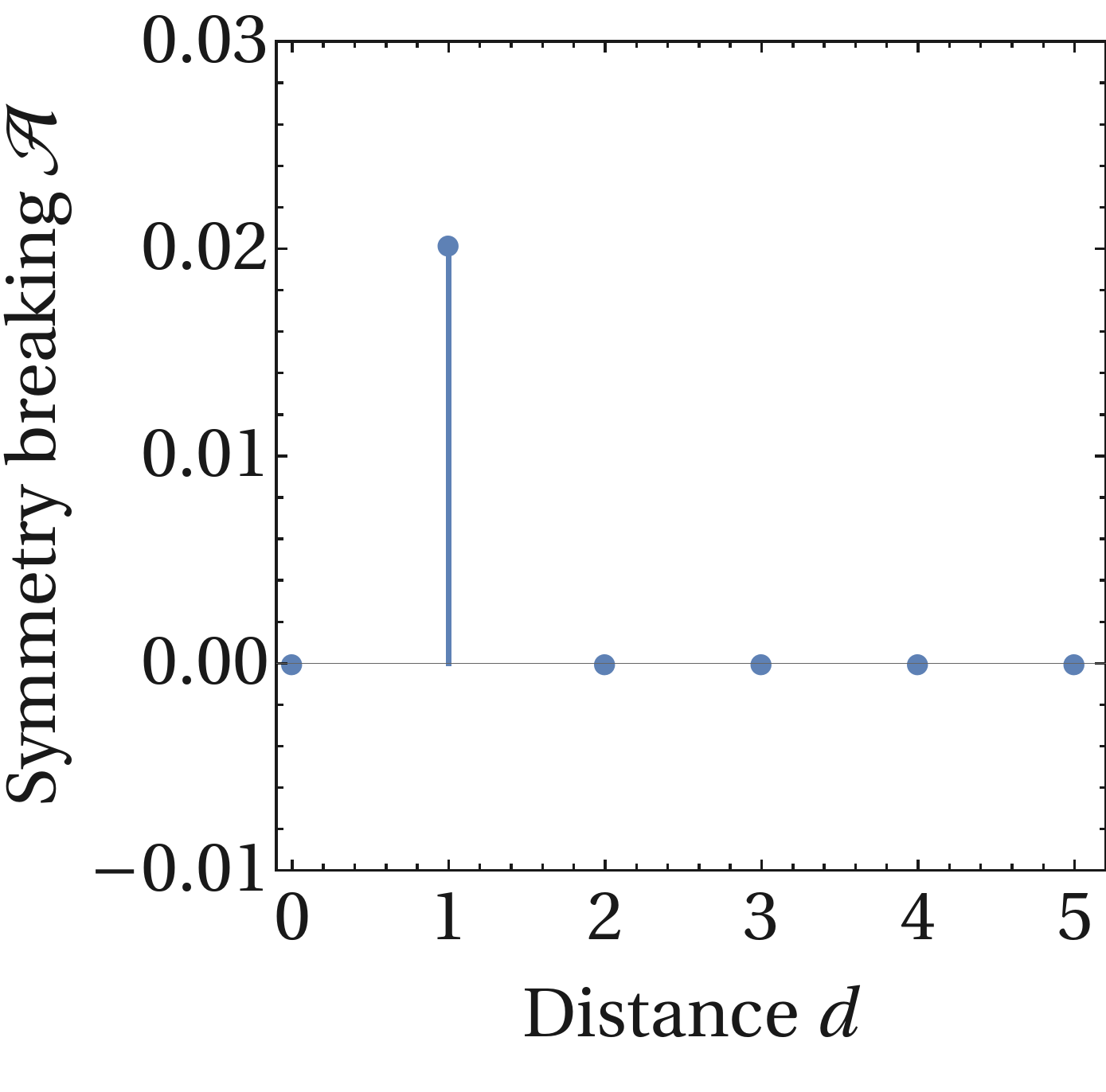}
\hspace{0.4cm}
\includegraphics[width=3.4cm]{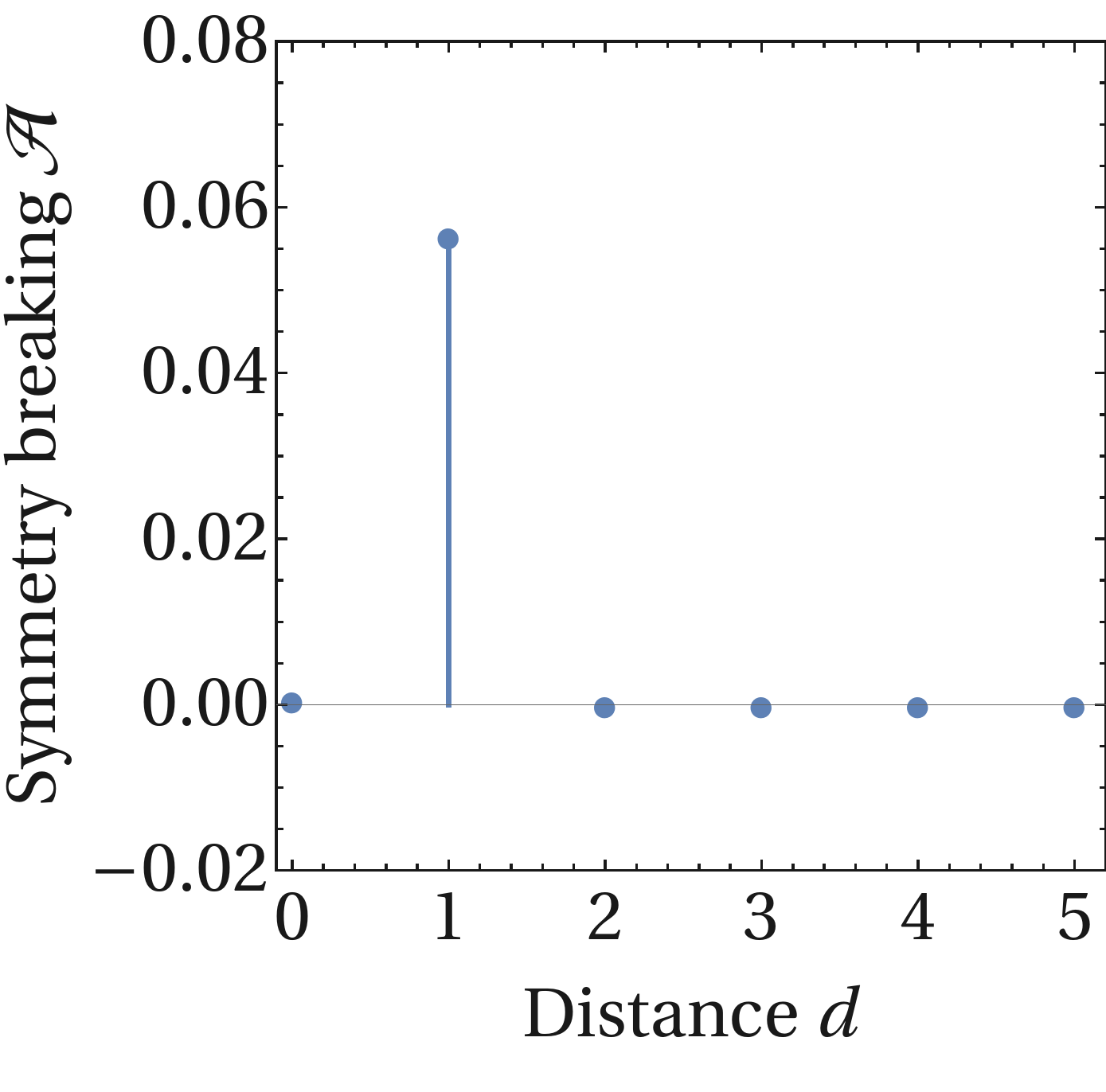}
\caption{$Z_2$ symmetry breaking at masses $m=10^{-1}$ (left) and $m=10^{-4}$ (right) in terms of asymmetry parameter $\mathcal{A}$ for block width $w=4$ and total size $N=60$.}\label{FIG_Z2_ASYM}
\end{figure}

As the $Z_2$ reflection symmetry is a property of the original system ($\rho_{AB}$) and leaves the EoP invariant, it might be natural to assume, as in \cite{Bhattacharyya:2018sbw}, that the optimal purification is $Z_2$-symmetric. However, we observe that to find the true minimum of $S_{A\tilde{A}}$ one needs to enlarge the parameter space by breaking the $Z_2$ exchange symmetry between $A\tilde A$ and $B\tilde B$. 
The results for $N=60$ are shown in Fig.\ \ref{FIG_EOP} (for the result assuming $Z_2$ symmetry, refer to Fig.\ \ref{FIG_EOP_Z2} in Appendix B). 
From $d=0$ to $d=1$, we observe a plateau-like behavior of $E_P$ at large $w$ whose width appears indendent of $w$, suggesting a finite-size effect. 
This notion is supported by the form of $K$ for minimal $S_{A\tilde{A}}$, shown in Fig.\ \ref{FIG_KMATRIX} for $m=10^{-4}$ and $w=4$. At $d=0$ and $d=1$, the couplings between $A_4 \leftrightarrow \tilde{B_1}$ and $\tilde{A}_4 \leftrightarrow B_1$ are enhanced, implying additional short-range entanglement.
The $Z_2$ symmetry breaking, while hard to discern from Fig.\ \ref{FIG_KMATRIX}, clearly appears at $d=1$ when considering the $Z_2$ asymmetry parameter $\mathcal{A}$ we defined in \eqref{EQ_Z2_ASYM}, shown in Fig.\ \ref{FIG_Z2_ASYM}. Within numerical accuracy, $\mathcal{A}=0$ for any $d \neq 1$. Note that the symmetry breaking becomes more pronounced with increasing $w.$ At $w=1$, it is not observable within numerical accuracy, while is clearly visible for the $w=4$ data in Fig.\ \ref{FIG_Z2_ASYM}.

\begin{figure}[tb]
\centering
\includegraphics[width=3.4cm]{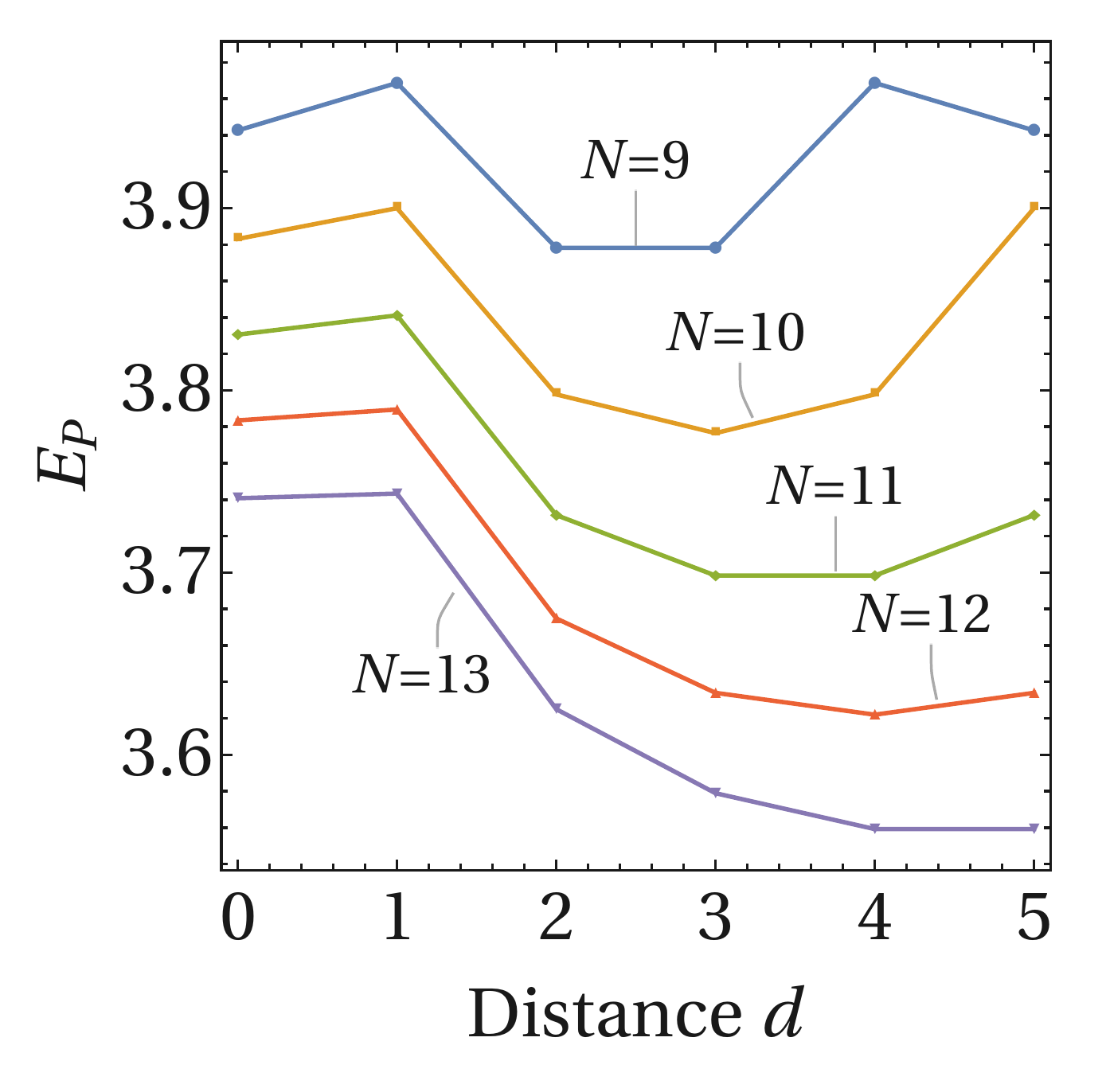}
\hspace{0.4cm}
\includegraphics[width=3.4cm]{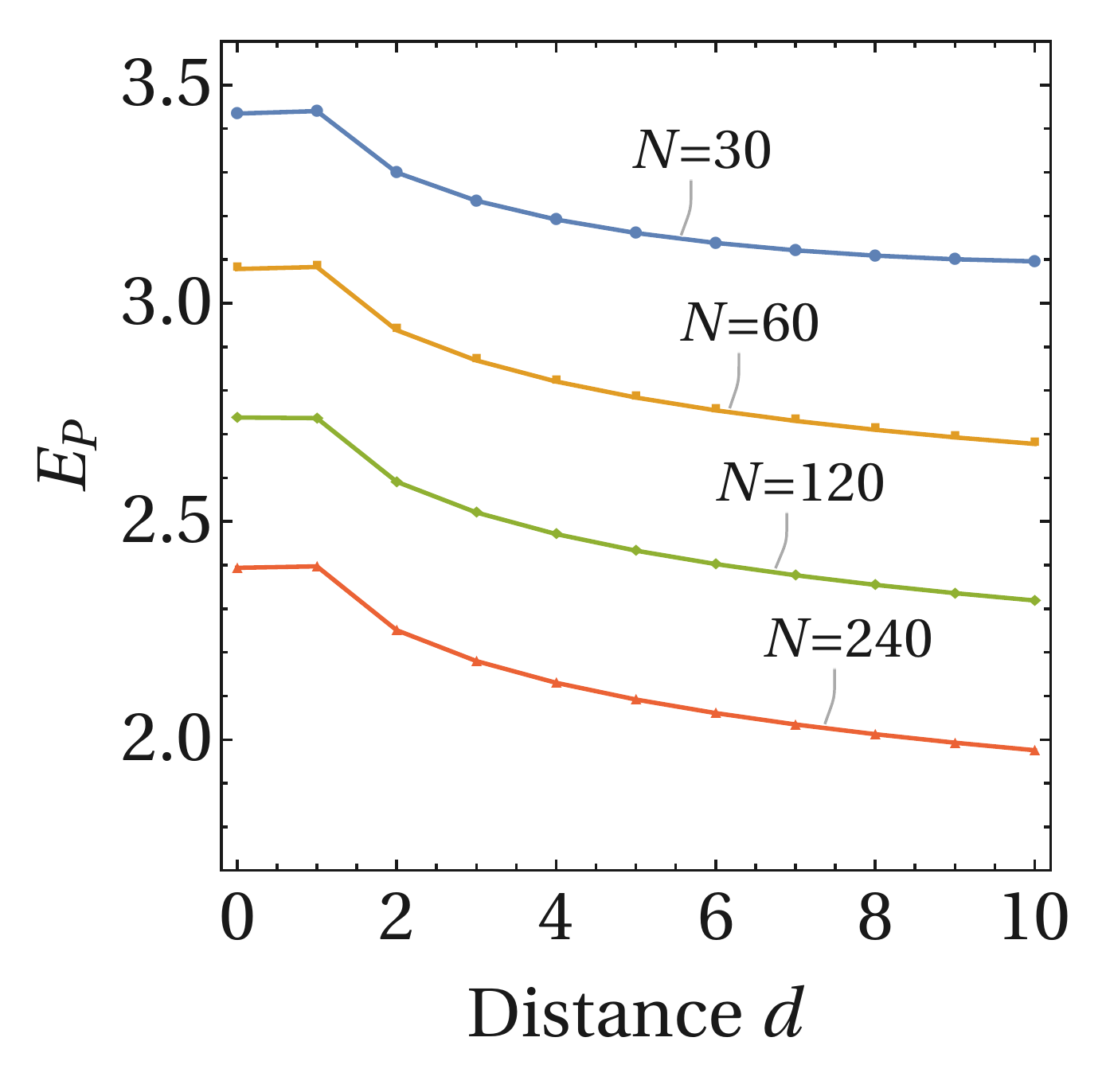}
\caption{EoP for small system sizes $N$ at block width $w=2$ and mass $m=10^{-4}$ (left) and for large sizes $N$ at $w=4$ and $m=10^{-4}$ (right).}\label{FIG_SMALL_LARGE_N}
\end{figure}

For small $N$,  the EoP does not monotonically decrease as a function of $d$ (Fig.\ \ref{FIG_SMALL_LARGE_N}, left). As we increase $N$, this non-monotonicity gradually disappears and leads to a plateau at large $w$ (Fig.\ \ref{FIG_SMALL_LARGE_N}, right). 
It is a surprise that the EoP, being a correlation measure, does not decrease monotonically with distance $d$ between $A$ and $B$, unlike the other correlation measures shown in Fig.~\ref{fig:NGB}. The same behavior can also appear in spin chains.\\

\begin{figure}[tb]
\centering
\includegraphics[width=3.4cm]{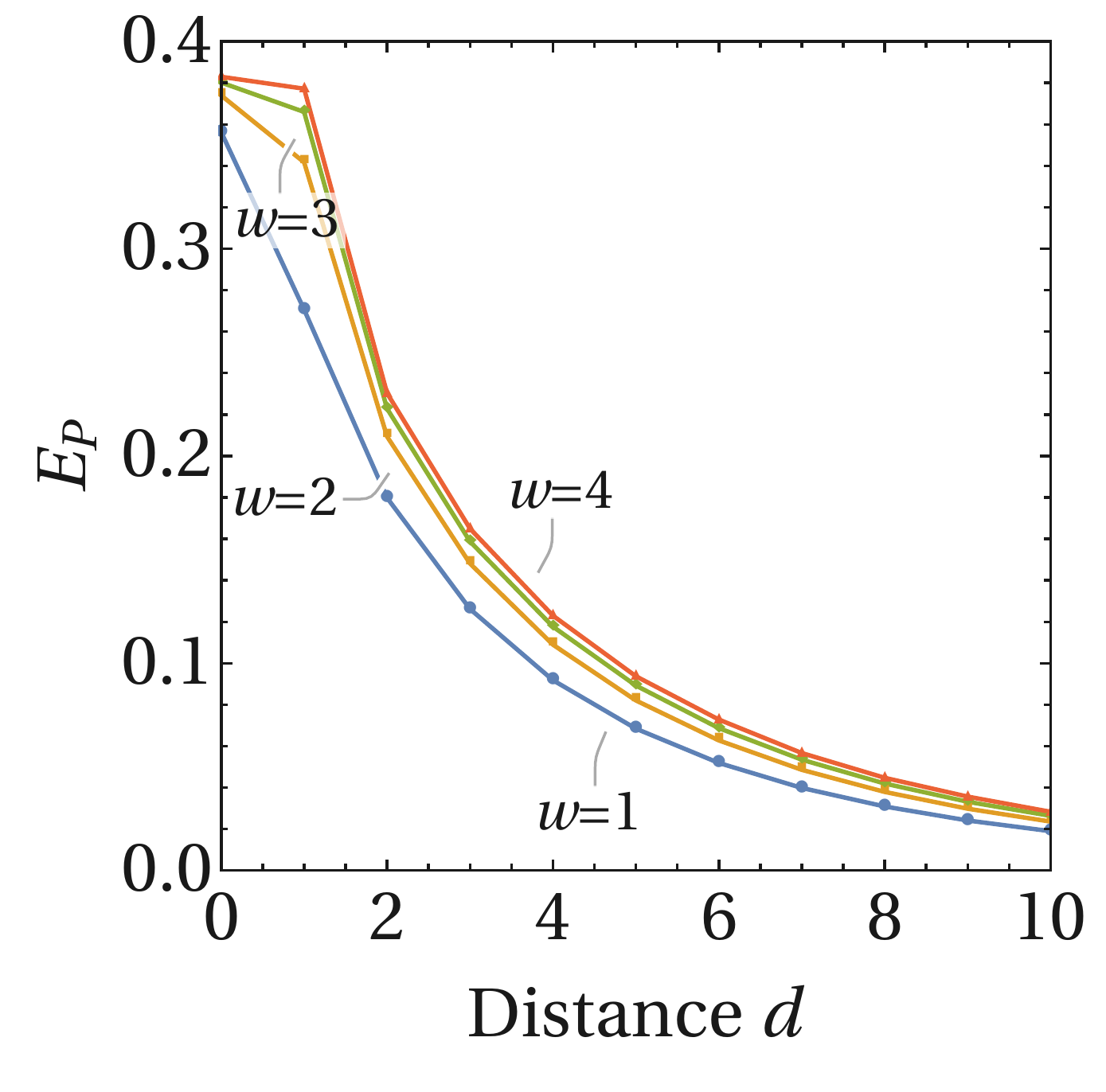}
\hspace{0.3cm}
\includegraphics[width=3.4cm]{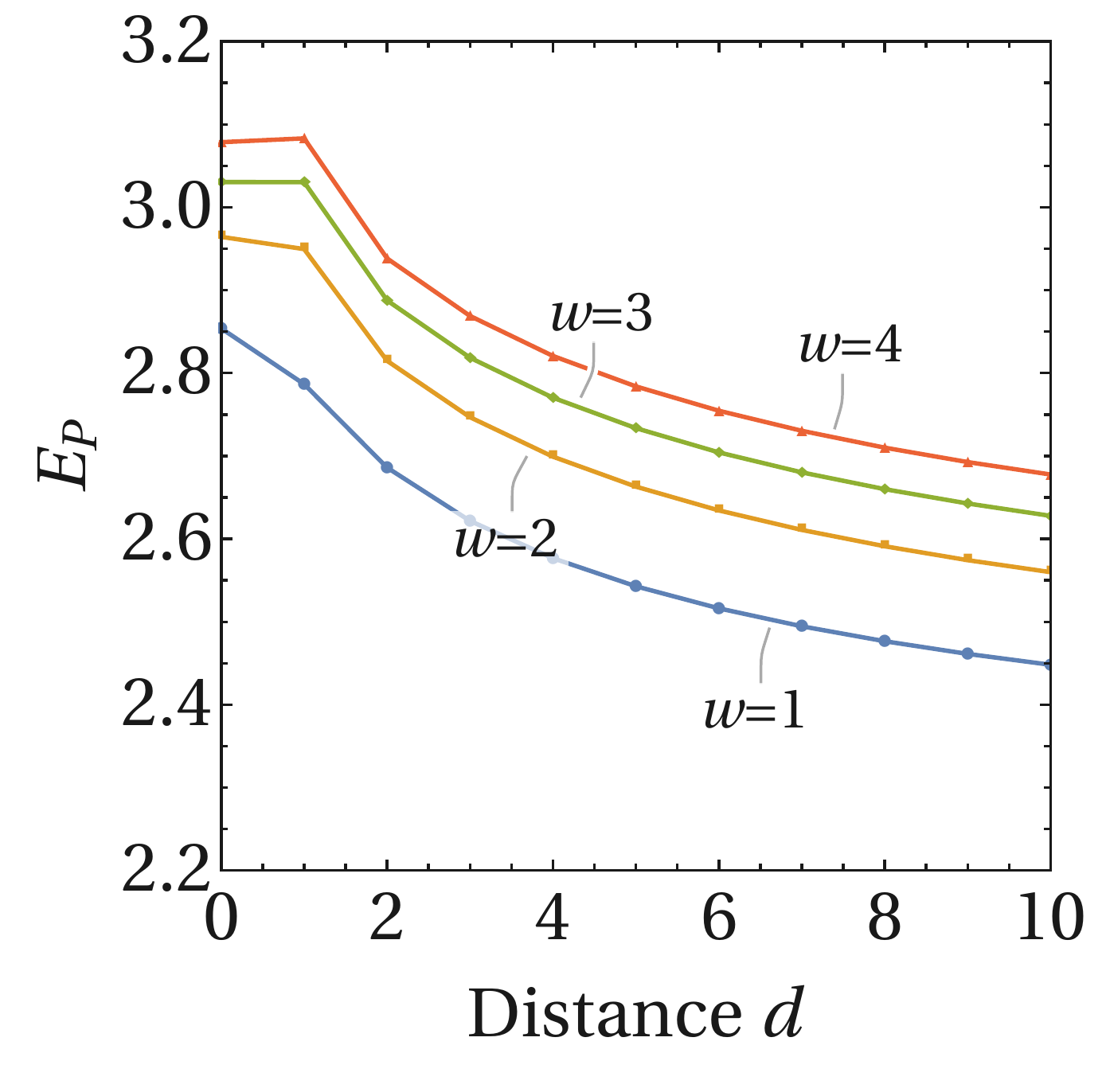}
\caption{Entanglement of purification $E_P$ at $N=60$ for $w=|A|=|B|=1,2,3,4$ (bottom to top) for mass $m=10^{-1}$ (left) and $m=10^{-4}$ (right), no $Z_2$ symmetry being assumed.
}
\label{FIG_EOP}
\end{figure}

{\bf 3. EoP in the transverse-field Ising model}

We will now compute the EoP for spin systems. 
Let us denote Hilbert space dimension by $D$ such that 
$D_A=\operatorname{dim}\mathcal{H}_A$ etc. 
In general, the dimension of auxiliary Hilbert space $D_{\ti{A}}D_{\ti{B}}$ should be at least as large as $\operatorname{rank}\rho_{AB}$ to purify a mixed state $\rho_{AB}$, with no general upper bound.  
Fortunately, the true minimum of $S_{A\ti{A}}$ can be found for
$D_{\ti{A}},D_{\ti{B}}\leq\operatorname{rank}\rho_{AB}$ in a system with finite-dimensional Hilbert space \cite{Robustness}, enabling us to compute EoP in practice.
For convenience, we call the purification  
{\it minimal}  when  $D_{\ti{A}}D_{\ti{B}}=\operatorname{rank}\rho_{AB}$, and 
{\it maximal} when $D_{\ti{A}}D_{\ti{B}}=(\operatorname{rank}\rho_{AB})^2$.
One  example of purification is the thermofield double purification (TFD)
\ba
\ket{\psi_{{\rm TFD}}}_{AB\ti{A}\ti{B}}=\sum_{i}\sqrt{p_{i}}\ket{i}_{AB}\ket{i}_{\ti{A}\ti{B}}\ , \label{TFD}
\ea
where 
we diagonalized the density matrix such that 
$\rho_{AB}=\sum_{i}p_{i}\ket{i}\bra{i}_{AB}$ 
with $\sum_{i}p_{i}=1$, $p_{i}\ge0$. 
The terminology {\it thermofield double} arises from the fact that the modular Hamiltonian $K_{AB}=-\log \rho_{AB}$ can be identified with the thermal Hamiltonian with inverse temperature $\beta=1$. 
All possible purifications of a fixed dimension can be obtained by acting with unitary operators on the auxiliary systems, yielding $\ket{\psi(U)}_{AB\ti{A}\ti{B}}=I_{AB}\otimes U_{\ti{A}\ti{B}}\ket{\psi_0}_{AB\ti{A}\ti{B}}$, where $\ket{\psi_0}_{AB\ti{A}\ti{B}}$ is an initial state. We also vary the dimensions $D_{\ti{A}}, D_{\ti{B}}$ to achieve both minimal and maximal purification. In principle, the maximal purifications are needed to obtain the EoP. However, we will find that often the minimal purification is sufficient to find the  true minimum of $S_{A\ti{A}}$.

We have used a variation of the steepest descent method, which is only guaranteed to converge to a \textsl{local} minimum of the objective function.
To obtain the global minimum, we start from several random initial purifications and ensure that the same point of convergence is reached. Nevertheless, the existence of additional local minima cannot be excluded, in which case the numerical results only provide an upper bound. The same is true for the scalar field case.

We deal with a 1D transverse-field Ising model
\begin{equation}
H_{{\rm Ising}}=-\sum_{\langle i,j \rangle}\sigma_{i}^{z}\otimes\sigma_{j}^{z}-h\sum_{i=1}^{N}\sigma_{i}^{x},
\end{equation}
where $\langle i,j \rangle$ denotes the summation over nearest neighbors with periodic
boundary condition and $N$ is the number of total spin sites. 

\begin{figure}
\centering
\includegraphics[width=3.4cm]{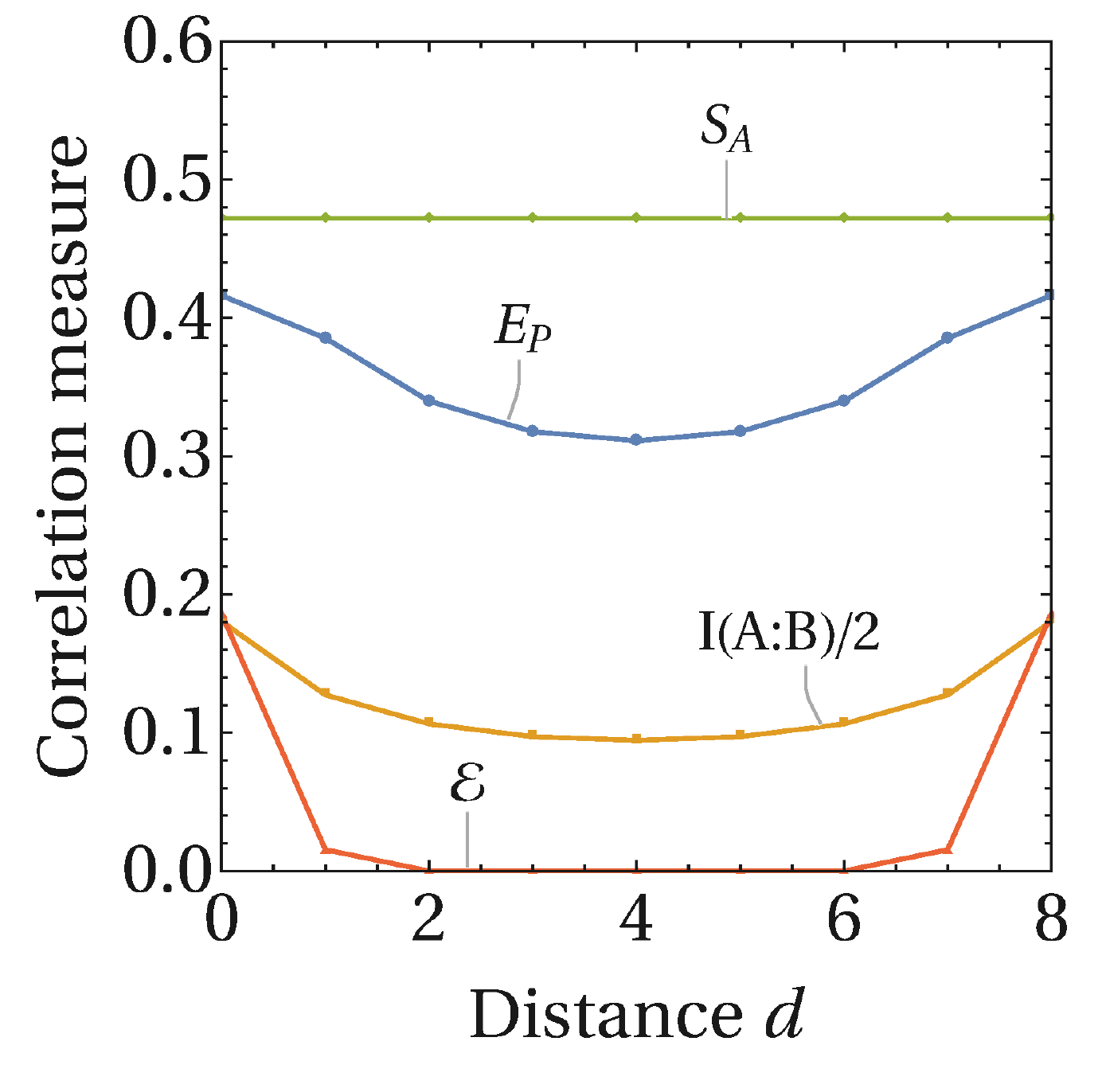}
\hspace{0.3cm}
\includegraphics[width=3.4cm]{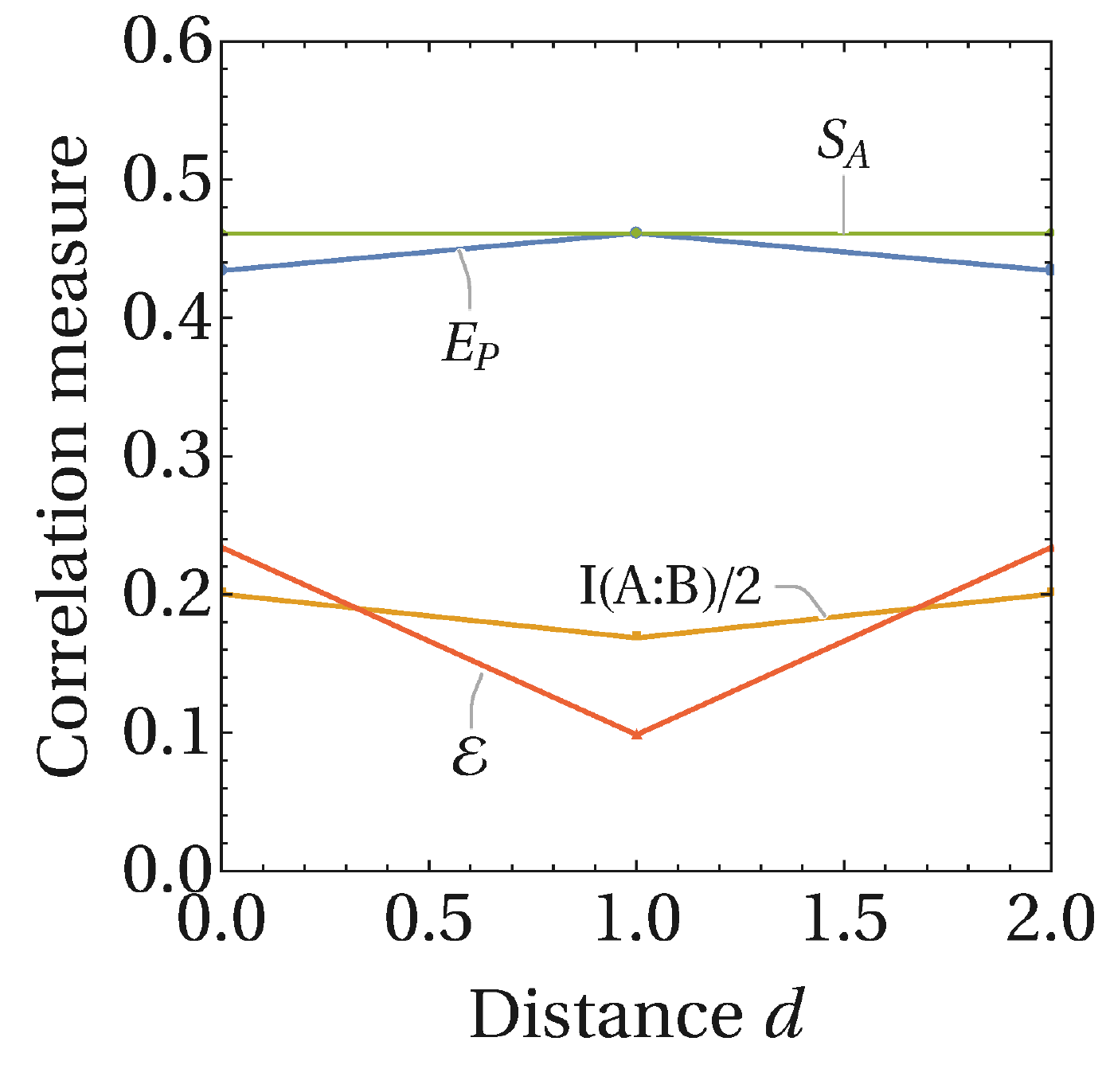}
\caption{\label{fig:EoPIsing} EoP for the critical Ising model at $w=1$, for $N=10$ (left) and $N=4$ (right).}
\end{figure}

First, we focus on the ground state of the critical Ising model ($h=1$) on $N$ sites.  The EoP for the corresponding subsystems with $w=|A|=|B|=1$ as a function of $d$ is depicted in Fig.\ \ref{fig:EoPIsing} along with MI and LN. 
While the optimization is performed for the maximal set of purifications, the optimal purification always corresponds to the minimal purification for this case.

\begin{figure}
\centering
\includegraphics[width=3.4cm]{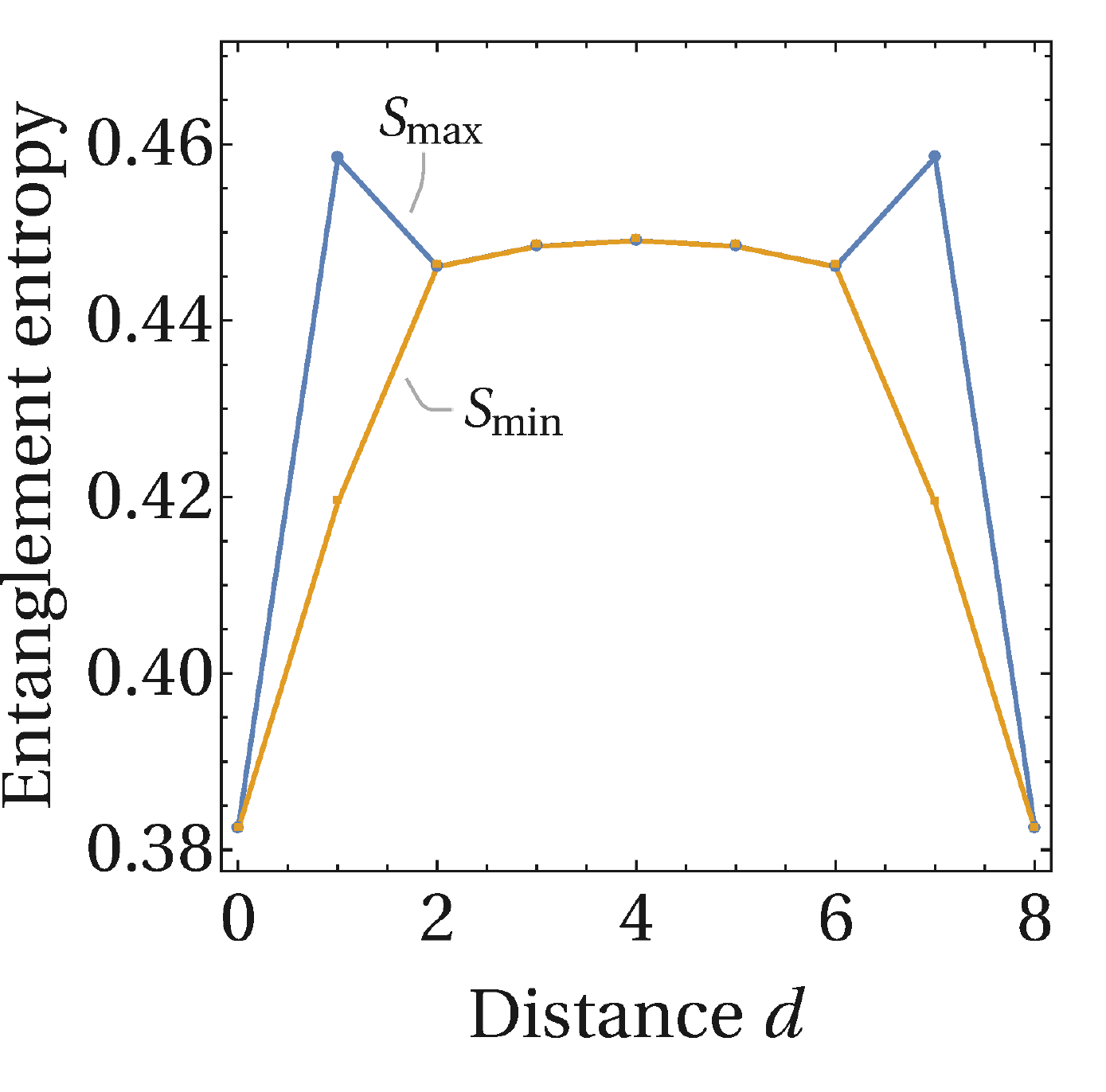}
\hspace{0.3cm}
\includegraphics[width=3.4cm]{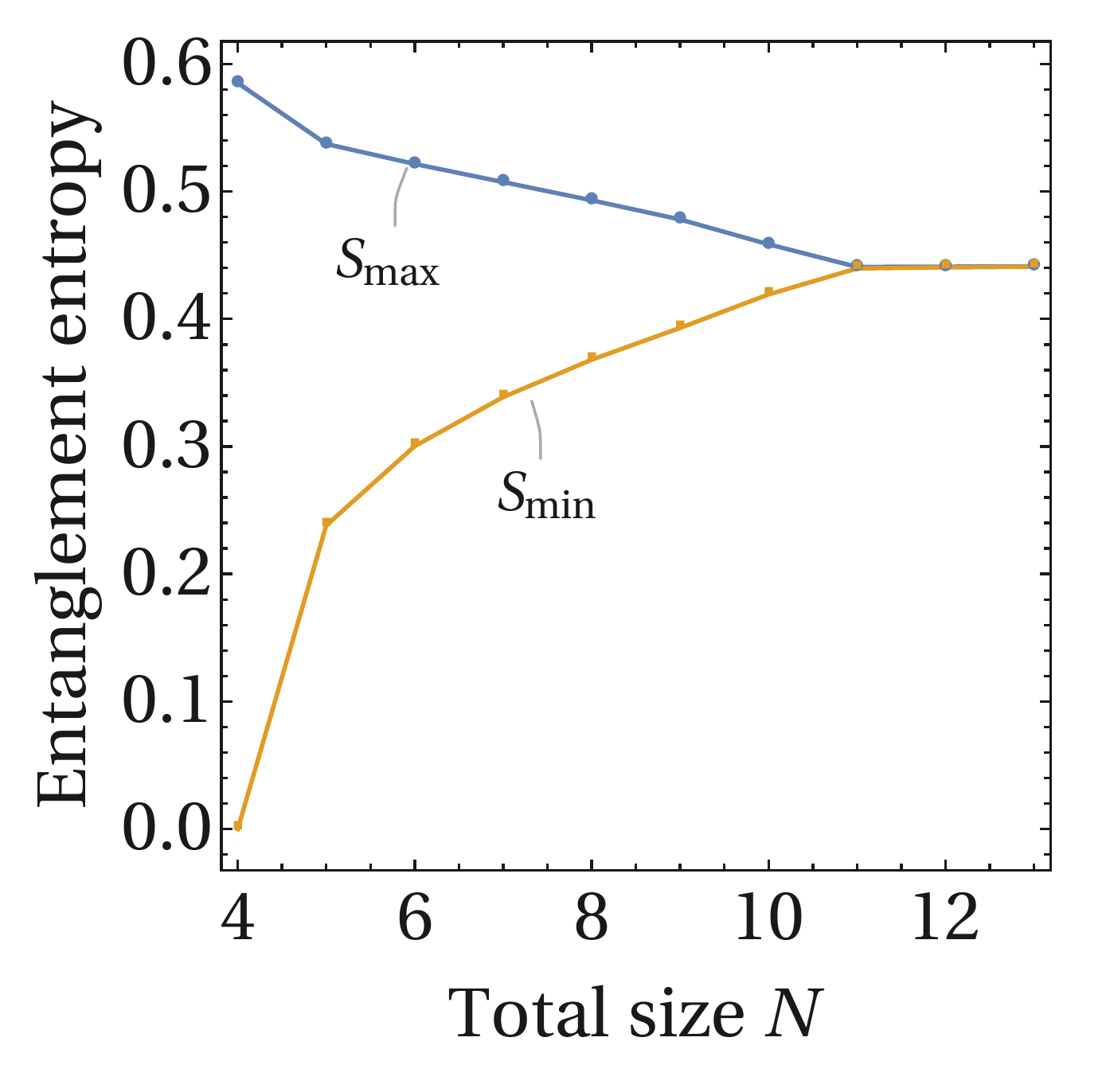}
\caption{\label{fig:EoPIsingSApSBp} $S_{\text{max}}=\operatorname{max} \{S_{\ti{A}},S_{\ti{B}}\}$ and $S_{\text{min}}=\operatorname{min} \{S_{\ti{A}},S_{\ti{B}}\}$ using the optimal purifications with $|A|=|B|=1$ in Ising model at $N=10$ (left). The $Z_2$ reflection symmetry of $\ti{A}$ and $\ti{B}$ at $d=1$ is recovered in the large $N$ limit (right).}
\end{figure}

For smaller $N$ ($N=4$), one can see that the EoP does not decrease with $d$. This can be explained as follows: $E_{P}$ must coincide with $S_A$ (Prop.\ 7 in \cite{Lock}) at $d=1$ since $\rho_{AB}$ has support only on a symmetric subspace, while $E_P<S_A$ at $d=0$ follows from the numerical computation (Fig.\ \ref{fig:EoPIsing}, right; see Appendix C for details).
This provides us with a clear example of EoP increasing with distance. 
Moreover, the $Z_2$ symmetry is clearly broken at $d=1$ as $S_{\ti{A}}$ $\neq S_{\ti{B}}$ (Fig.\ \ref{fig:EoPIsingSApSBp}). As in the scalar case, the $Z_2$ symmetry breaking leads to two degenerate configurations for $E_p$, related by an $A\ti{A} \leftrightarrow B \ti{B}$ flip. 
Moreover, $S_A \neq S_{\ti{A}}$ implies that the optimal purifications are not TFD purification.
For $w=1$, the $Z_2$ symmetry is gradually recovered as $N$ gets larger ($N \gtrsim 12$).

\begin{figure}
\centering
\includegraphics[width=3.4cm]{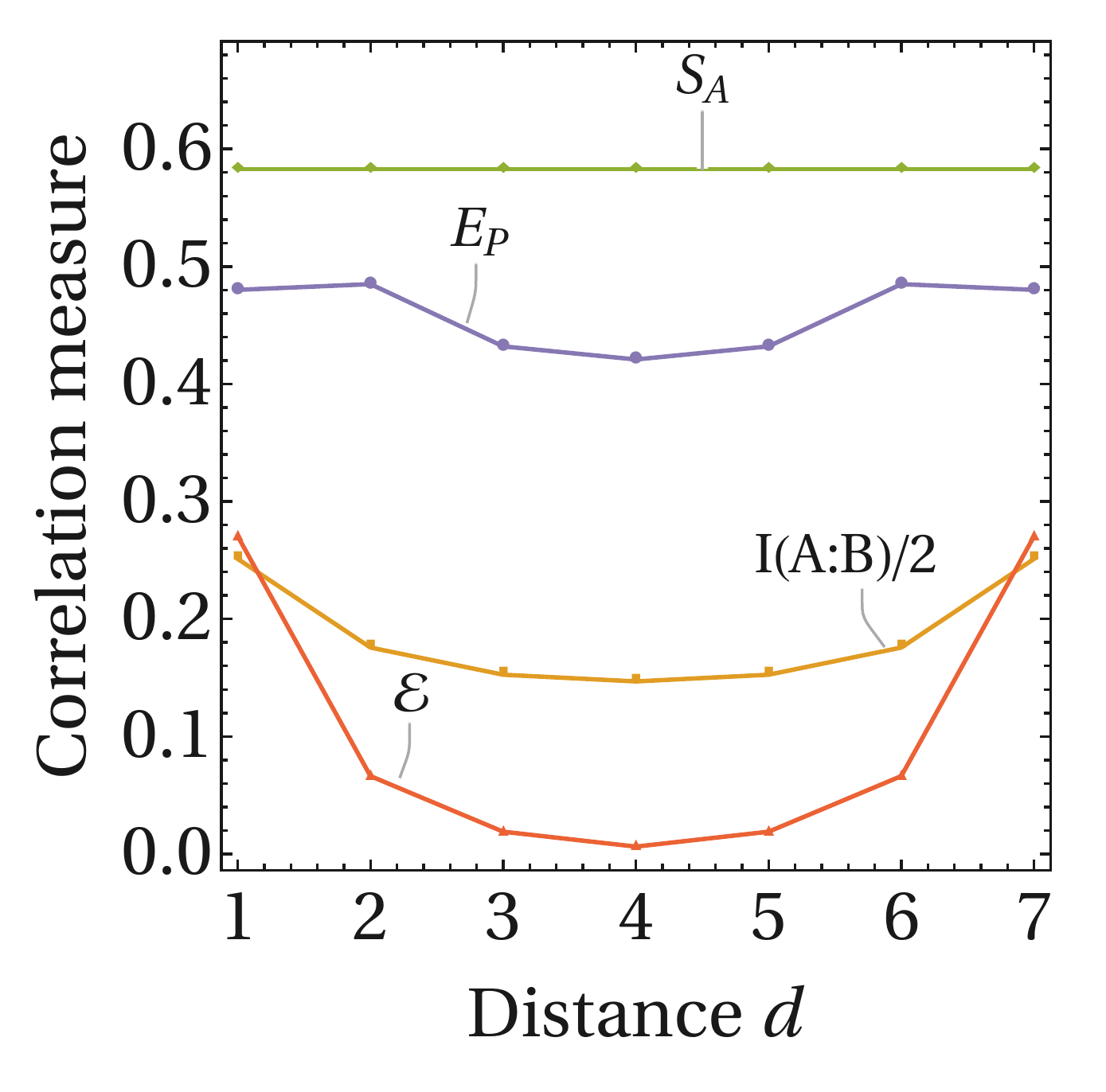}
\hspace{0.3cm}
\includegraphics[width=3.4cm]{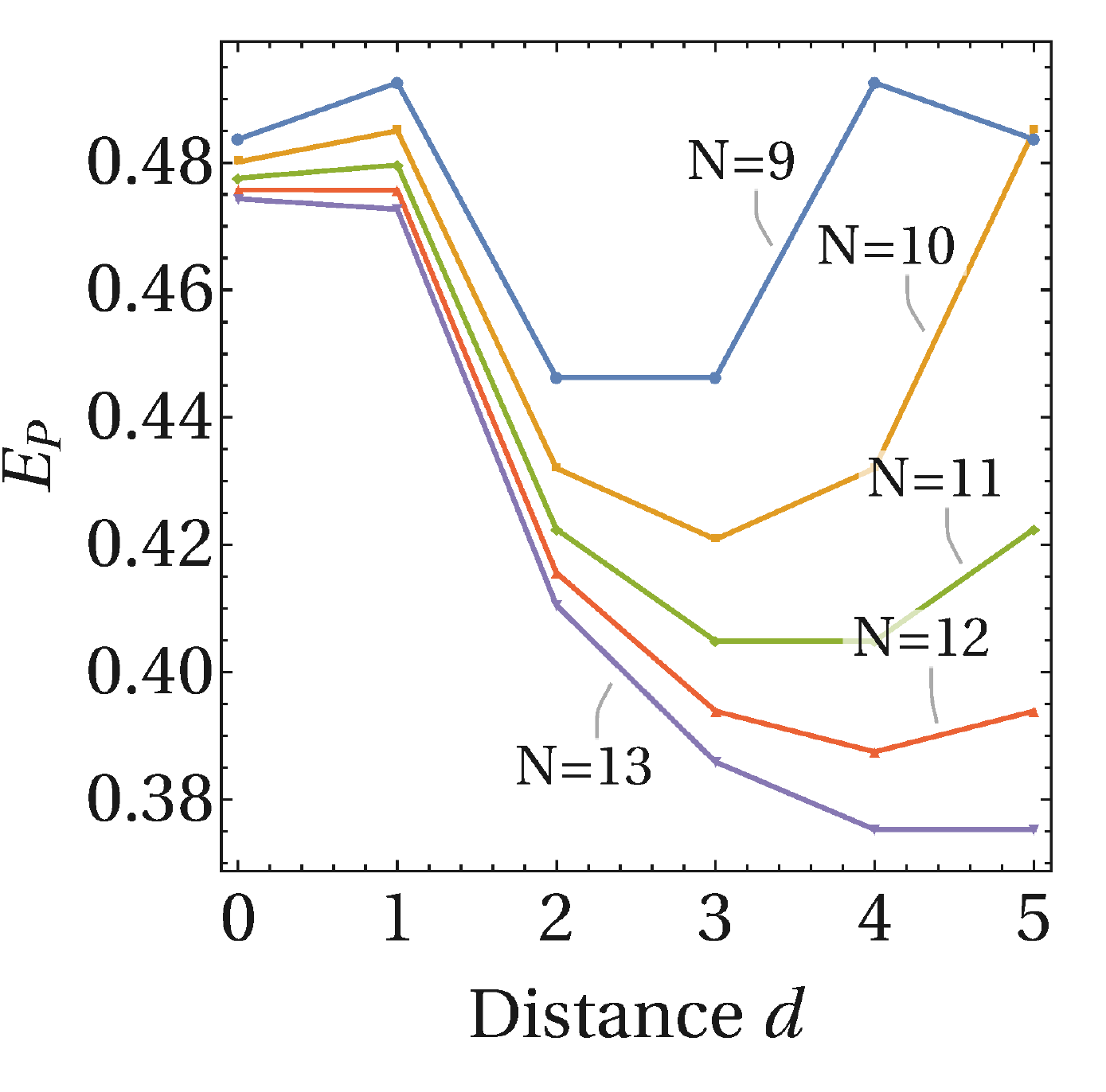}
\caption{\label{fig:EoPIsingA2B2} EoP for the critical Ising model at $w=2$, for $N=10$ (left) and various $N$ (right). The non-monotonic behavior gets weaker as $N$ increases. All results for $w=2$ are optimized within minimal purifications.}
\end{figure}

We also consider the larger subsystem size $w=2$. In this case the EoP is computed using minimal purifications to expedite the computation. We again observe a non-monotonic behavior of EoP that weakens as $N$ increases (Fig.~\ref{fig:EoPIsingA2B2}), similar to the free scalar case.
The $Z_2$ symmetry breaking is also found at $d=1$, which remains even at large $N$. 

Both a plateau and a $Z_2$ symmetry breaking occur also for a class of two-qubit states called \textsl{Werner states}, which coincide with the ground state of the Heisenberg spin chain. For details, refer to Appendix D.\\

{\bf 4. Phase transition and $Z_2$ symmetry breaking in the Ising model}

\begin{figure}
\centering
\includegraphics[width=3.4cm]{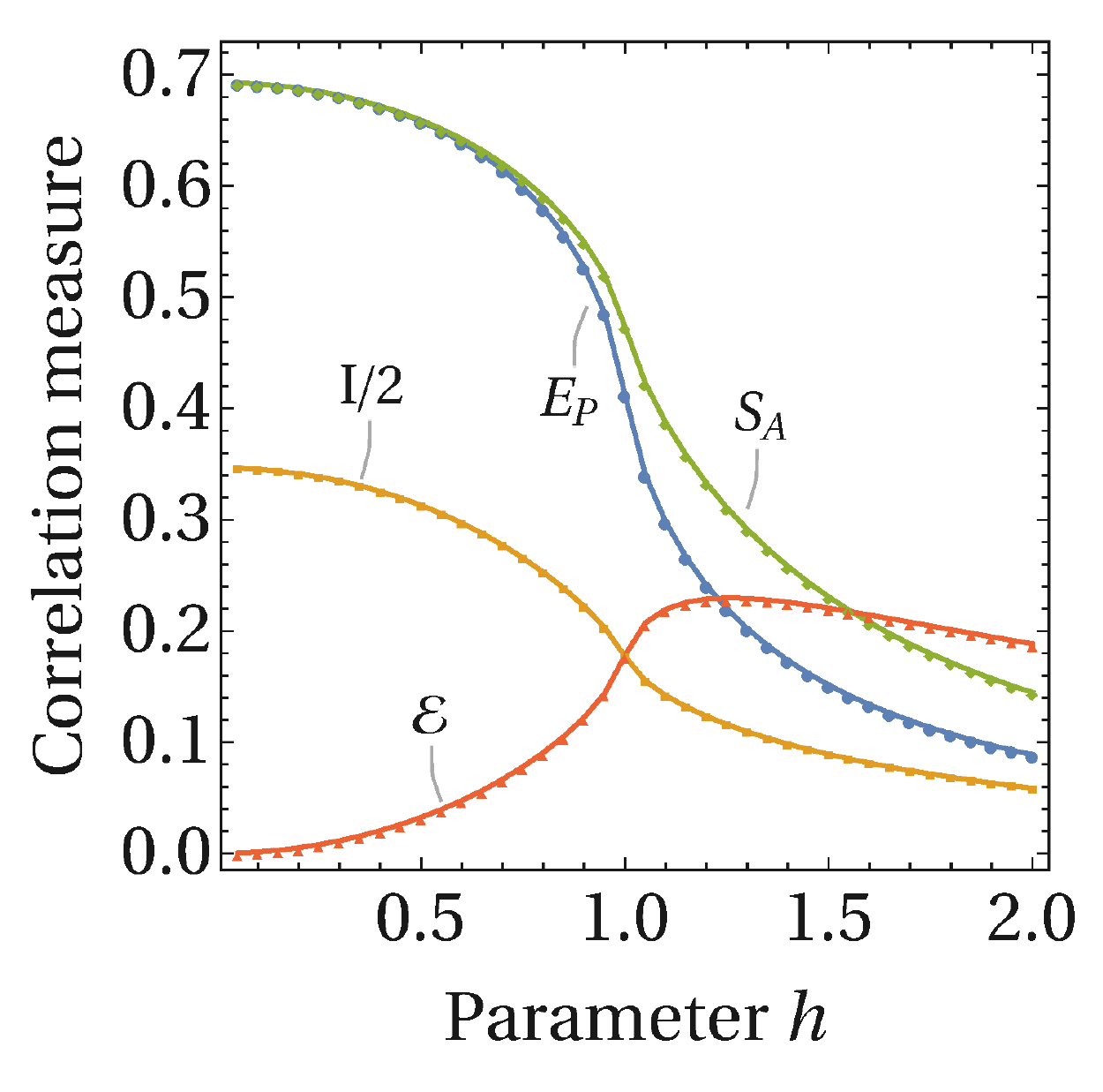}
\hspace{0.3cm}
\includegraphics[width=3.4cm]{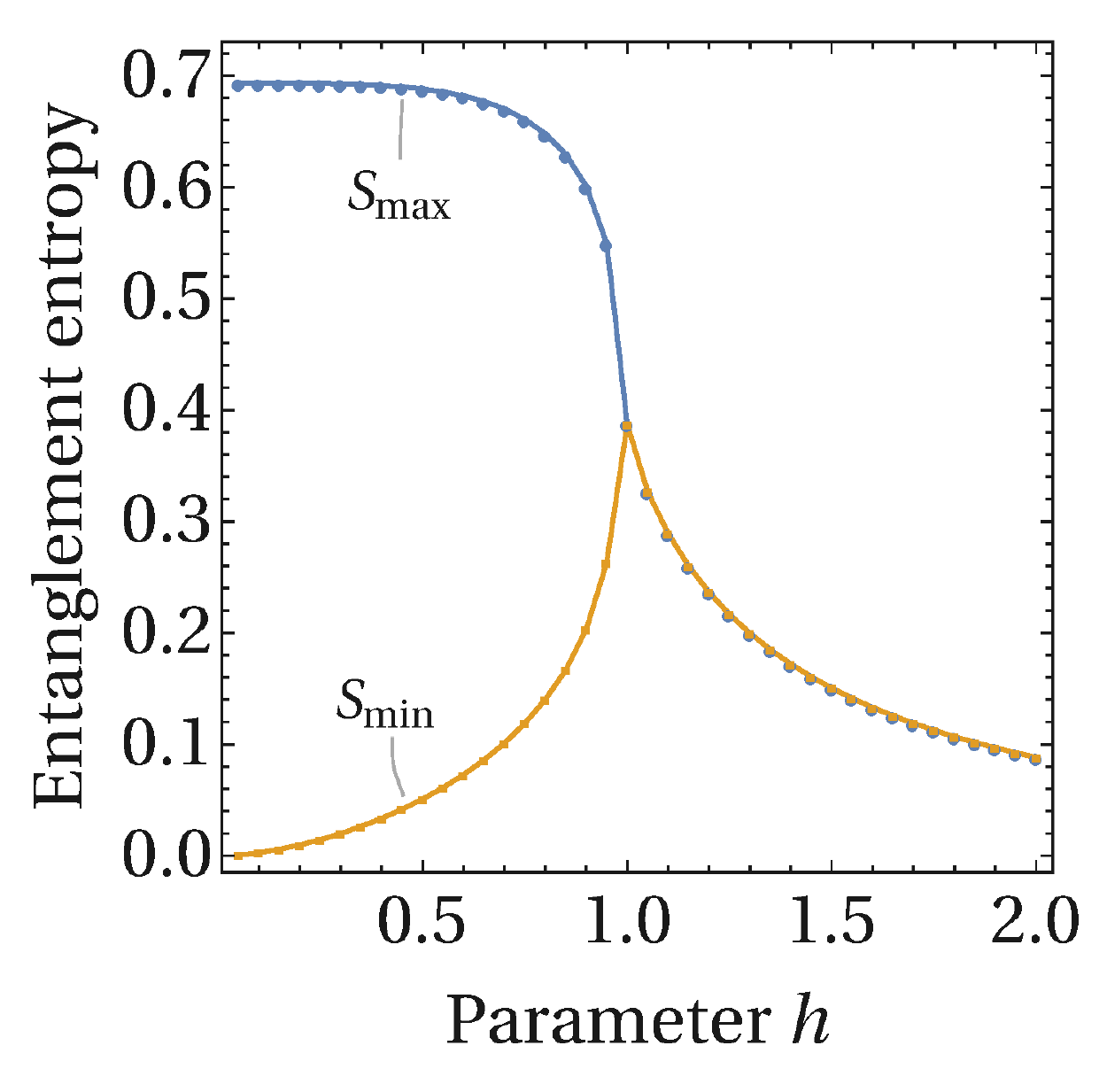}
\caption{\label{fig:EoPIsingPT} EoP in the transverse-field Ising model with a magnetic field $h$ for $d=0$, $w=1$, and $N \rightarrow \infty$ (left). The $Z_2$ symmetry is broken in the ferromagnetic phase $h<1$,  resulting in different entanglement entropies $S_{\text{max}}{=}\operatorname{max} \{S_{\ti{A}},S_{\ti{B}}\}$ and $S_{\text{min}}{=}\operatorname{min} \{S_{\ti{A}},S_{\ti{B}}\}$ (right).}
\end{figure}

Furthermore, we compute the EoP as a function of the magnetic field $h$ for the nearest-neighbor minimum subsystems in the thermodynamic limit $N \to \infty$. We consider the whole system to be  in the thermal ground state for which the analytic form of the reduced density matrices is obtained \cite{EEIsing, IsingSolved}. The result in Fig.~\ref{fig:EoPIsingPT} shows that the EoP has an inflection point at $h=1$. It indicates that the EoP correctly captures the phase transition of the original system. Remarkably, the $Z_2$ reflection symmetry of $A\ti{A}$ and $B\ti{B}$ gets broken only in the ferromagnetic phase $h<1$. However, the thermal ground state maintains a $Z_2$ flip symmetry in any phase. \\
This may imply a connection between the physical phase transition and $Z_2$ reflection symmetry breaking.\\ 

{\bf 5. Conclusions and Discussion}

Finally, we seek to provide an interpretation of our results.
For both free scalar theory and the critical Ising model,
we observed a non-monotonic or plateau-like behavior of the EoP at small $d$. 
These behaviors are very special to EoP and do not appear in MI. This is in contrast to the fact that they possess similar information-theoretic properties as total correlation measures (refer to e.g.\ \cite{BP}). 
This mirrors the observation in \cite{UT,Nguyen:2017yqw} that the value of holographic EoP behaves differently than that of holographic MI, with the former developing a plateau-like behavior.

Suppose total correlations (measured by half of MI) are a combination of quantum entanglement and classical correlations. As $E_{P}\geq 2(I(A:B)/2)$ for separable states \cite{EP} while $E_P=I/2$ for pure states, we assert that EoP enhances the classical correlations compared to $I(A:B)/2$ at least twofold, while treating quantum entanglement equivalently.
This explains the non-monotonicity  of EoP as well: Quantum entanglement can be estimated by the LN, which falls of quickly with $d$. Thus classical contributions at $d\geq 1$ are enhanced compared to short-range quantum entanglement at $d=0$. 
Possible connections to analogous quantities such as quantum discord \cite{Discord1,Discord2} will be an interesting future work.\par

\begin{figure}[tt]
  \centering
  \includegraphics[width=2.42cm]{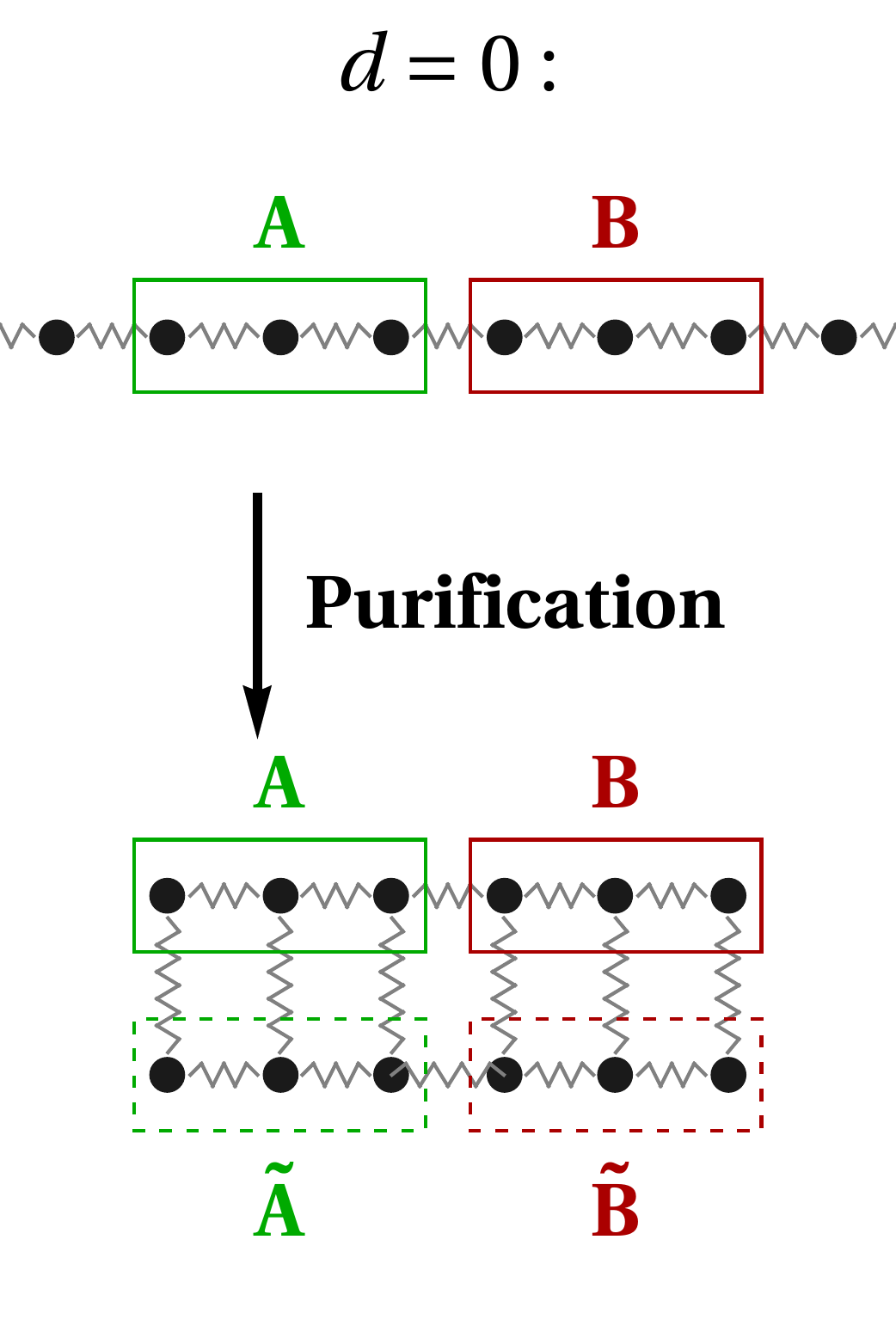}
  \hspace{0.4cm}
  \includegraphics[width=2.42cm]{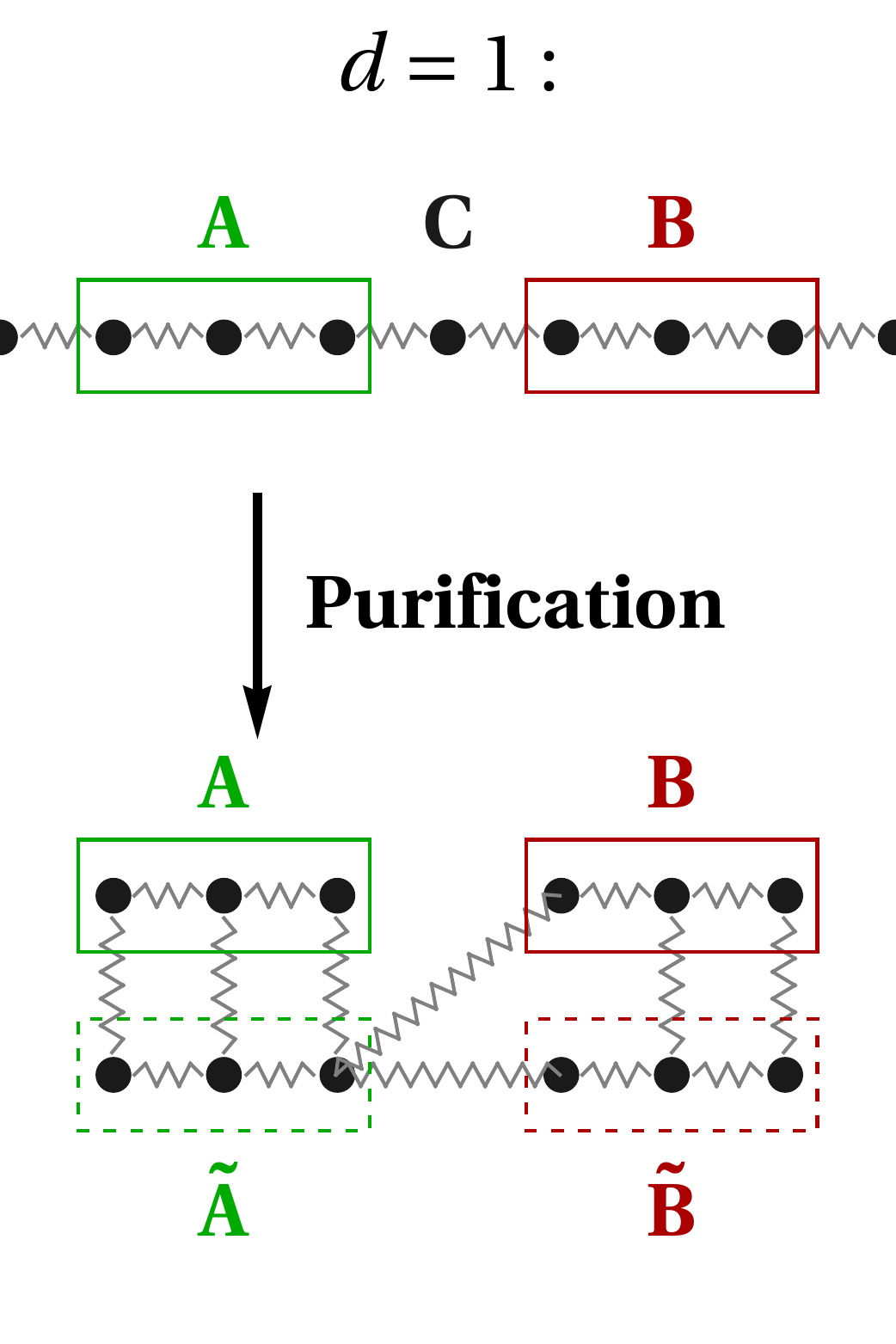}
  \hspace{0.4cm}
  \includegraphics[width=2.42cm]{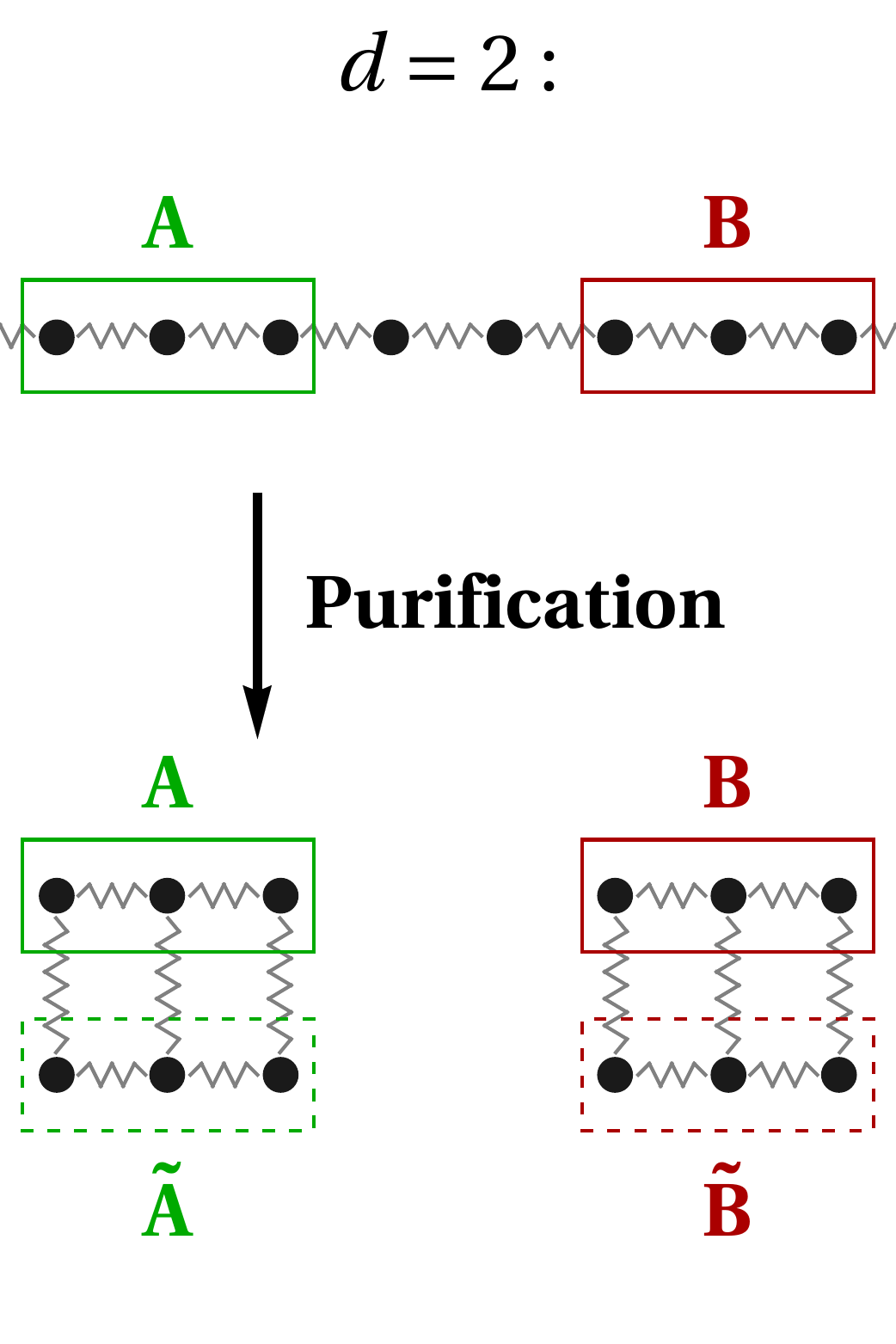}
  \caption{A toy model for EoP for 1D many body systems, assuming only short-range quantum entanglement (zigzag lines) for $w{=}|\ti{A}|{=}|\ti{B}|{=}3$.
  At $d=1$, we only show one of the two optimal $Z_2$ symmetry-broken purifications.} 
\label{fig:setupt}
  \end{figure}
  
We also propose a mechanism of $Z_2$ symmetry breaking at $d=1$ by a toy model with dominant nearest-neighbor quantum entanglement (Fig.\ \ref{fig:setupt}). The distinction between quantum entanglement and classical correlation is crucial here, as well.
At $d=1$, an intermediate site $C$ is strongly entangled with both $A$ and $B$, and tracing it out turns $\rho_{AB}$ into a highly mixed state. This leads to strong classical correlations between $A$ and $B$. 
As a result, the purification requires strong entanglement for $A\leftrightarrow\ti{A}$,  $B\leftrightarrow\ti{B}$, $\ti{A}\leftrightarrow\ti{B}$, $\ti{A}\leftrightarrow B$ and $A\leftrightarrow \ti{B}$ in order to convert the large amount of classical correlations into quantum entanglement. 
This complicated competition, under the constraint of monogamy, results in a $Z_2$ reflection symmetry breaking, where only either $\ti{A}\leftrightarrow B$ or $A\leftrightarrow \ti{B}$ exhibit strong entanglement (Fig.\ \ref{fig:setupt}, center). This picture is indeed confirmed both for the free scalar and the Ising model. 
In contrast, correlations are either weak at $d\geq2$ or are strong but mainly consist of entanglement at $d=0$. Both cases require little purification, allowing a simple symmetric purification to be optimal. This suggests that the $Z_2$ symmetry breaking occurs when $\rho_{AB}$ posseses strong classical correlations.\par
Notice that the $Z_2$ symmetry breaking does not occur for CFT vacua in holographic setups. However, such a symmetry breaking can be possible in holography for excited states or non-conformal setups. 
Searching for $Z_2$ symmetry breaking in holographic EoP will thus serve as an interesting future endeavor. 

In our analysis of the EoP for the transverse-field Ising model, we found that the $Z_2$-broken region coincides with the ferromagnetic phase. This suggests an interesting connection between symmetry breaking in the optimal purification for the EoP and a quantum phase transition. This deserves future studies.\\

\begin{acknowledgments}
\section*{Acknowledgments}
We thank Pawel Caputa, Horacio Casini, Jens Eisert, Masamichi Miyaji, Masahiro Nozaki, Kazuma Shimizu, and Brian Swingle for useful conversations. We are very grateful to Yoshifumi Nakata for valuable comments on the draft. 
AB and KU are supported by JSPS fellowships. AJ is supported by a \textsl{Studienstiftung} fellowship. TT is supported by the Simons Foundation through 
the ``It from Qubit'' collaboration. TT is supported by JSPS Grant-in-Aid for Scientific Research (A) No.16H02182 
and by JSPS Grant-in-Aid for Challenging Research (Exploratory) 18K18766. AB and TT are supported 
by JSPS Grant-in-Aid for JSPS fellows 17F17023.
KU is supported by Grant-in-Aid for JSPS Fellows No.18J22888. 
TT is also supported by World Premier International Research Center Initiative 
(WPI Initiative) from the Japan Ministry of Education, Culture, Sports, Science and Technology (MEXT).
\end{acknowledgments}

\section{Appendix A: Computing Negativity in Free Scalar Field Theory}\label{app:NG}

One simple characterization of quantum entanglement between subsystems $A$ and $B$ for a mixed state 
$\rho_{AB}$ is the logarithmic negativity \cite{VW}. For this we introduce the so-called partial transposition $\Gamma_B$, 
which is the transposition acting only for the subsystem $B$. It is well-known that
for separable states, the partially transposed density matrix $\rho^{\Gamma_B}_{AB}$ is still positive,
while for non-separable (=entangled) states , this positivity is not preserved in general.
We would like to note that even if $\rho^{\Gamma_B}_{AB}$ is positive, we cannot say $AB$ is separable, while the converse statement is true.

The logarithmic negativity is defined by
\be
\mathcal{E}_{N}(\rho_{AB})=\log \Tr |\rho_{AB}^{\Gamma_{B}}|,
\ee
where we introduced
\be
\Tr |\rho_{AB}^{\Gamma_{B}}|=\Tr\sqrt{(\rho_{AB}^{\Gamma_{B}})^{\dagger} \rho_{AB}^{\Gamma_{B}}}.
\ee

If we write the eigenvalues of $\rho^{\Gamma_B}_{AB}$ as $\lambda_i$, then we can write
\be
\mathcal{E}_{N}(\rho_{AB})=\log \left(\sum_i|\lambda_i|\right)\geq 0.
\ee
Note here that since  $\rho_{AB}$ and  $\rho^{\Gamma_B}_{AB}$ are both normalized, we have
$\sum_i\lambda=1$ and thus $\sum_i|\lambda|>1$. The logarithmic negativity is vanishing if and only if
all the eigenvalues $\lambda_i$ are non-negative. This quantity is known to be monotonic under LOCC and satisfies at least the minimal property of an entanglement measure for mixed states. Also note that when the total state $\rho_{AB}$ is pure, $\mathcal{E}_{N}(\rho_{AB})$ is not equal to the (von Neumann) entanglement entropy but is equal to the $n=1/2$ R\'{e}nyi entropy, defined by
$S^{(1/2)}_A=2\log \Tr (\rho_A)^{1/2}$.

Now we compute logarithmic negativity  for the ground state $\Psi_0$ for our free scalar lattice model. We divide the total lattice system into subregions $A, B$ and $C$ such that ${\cal H}_{tot}={\cal H}_A\otimes {\cal H}_B\otimes {\cal H}_C$. We define their lattice sizes to be $|A|,|B|$ and $|C|$.
In this setup, we wish to compute the logarithmic negativity which measures the quantum entanglement between $A$ and $B$.
First remember that the ground state wave functional is given by (\ref{EQ_PHI0_B}). Then the reduced density 
matrix $\rho_{AB}=\mbox{Tr}_{C}
\left[|\Psi_{ABC}\lb \la\Psi_{ABC}|\right]$  is obtained by integrating out $C$:
\ba
&& \rho_{AB}[\phi_{AB},\phi'_{AB}]\no
&&=\int D\phi_C \Psi^*_0[\phi_{AB},\phi_C]\cdot\Psi_0[\phi'_{AB},\phi_C]\no
&&\propto \exp\Biggl[-\frac{1}{2}(\phi_{A},\phi_{B}) M \left(\begin{array}{c}
  \phi_{A}\\
  \phi_{B}
\end{array}\right)-\frac{1}{2}(\phi'_{A},\phi'_{B}) M \left(\begin{array}{c}
  \phi'_{A}\\
  \phi'_{B}
\end{array}\right)\no
&&\ \ \ \ \ \ \ -\frac{1}{4}\Big(\phi_A-\phi'_A,\phi_B-\phi'_B\Big)
N \left(\begin{array}{c}
  \phi_{A}-\phi'_A \\
  \phi_{B}-\phi'_B
\end{array}\right)
\Biggr], \nonumber \\\label{rab}
\ea
where $M$ and $N$ are symmetric real matrices.  They are defined as
\ba
M=P-Q R^{-1} Q^{T}, N=Q R^{-1} Q^{T}.
\ea
For later purpose, it is useful to decompose $M$ and $N$, which are $(|A|+|B|)\times (|A|+|B|)$ matrices, into $|A|\times |A|$,$|A|\times |B|$ and $|B|\times |B|$ matrices as follows:
\be
M= \left(
   \begin{array}{cc}
     M_1 & M_2 \\
     M^T_2 & M_3 \\
   \end{array}
 \right),\ \ \ \
  N= \left(
   \begin{array}{cc}
     N_1 & N_2 \\
     N^T_2 & N_3 \\
   \end{array}
 \right),
\ee
where ${}^T$ is the standard transposition.

Given this density matrix we now proceed to compute negativity. For that, we have to first perform the partial transpose $\Gamma_B$,  which is equivalent to
interchanging $\phi_{B}$ and $\phi'_B.$
After we rearrange this as a matrix whose arguments are of the form $(\phi_{A},\phi_{B}),(\phi'_{A},\phi'_{B})$, we obtain:
\ba
&& \rho^{\Gamma_{B}}_{AB}[(\phi_{A},\phi_{B}),(\phi'_{A},\phi'_{B})]\no
&&\propto \exp\Biggl[-\frac{1}{2}(\!\phi_{A},\!\phi_{B}) M \left(\begin{array}{c}
  \!\phi_{A}\\
  \!\phi_{B}
\end{array}\right)\!-\frac{1}{2}(\!\phi'_{A},\!\phi'_{B}) M \left(\begin{array}{c}
 \! \phi'_{A}\\
 \! \phi'_{B}
\end{array}\right)\no
&&\  \ \ \ \ \ \ \ -\frac{1}{4}\Big(\phi_A\!-\!\phi'_A,\phi_B\!-\!\phi'_B\Big)
\tilde N \left(\begin{array}{c}
  \!\phi_{A}-\phi'_A \\
  \!\phi_{B}-\phi'_B
\end{array}\right)
\Biggr], \nonumber\\\ea
where
\ba
\tilde N= \left(\begin{array}{cc}
 N_1 & -N_2-2M_2 \\
  -N^T_2-2M^T_2 & N_3
\end{array}\right).
\ea

Now we can perform a field redefinition:
\be
\hat \phi_{AB}= M_{D}^{1/2} V \phi_{AB},
\ee
where $V$ is a orthogonal matrix and $M_D$ is a diagonal matrix.
We choose them such that we have
\be
 M= V^{T} M_{D} V.
\ee
We apply the same transformation on $\phi'_{AB}.$
Then we find
\ba
&& \rho^{\Gamma_{B}}_{AB}[(\hat \phi_{A},\hat \phi_{B}),(\hat\phi'_{A},\hat\phi'_{B})]\no
&&\propto \exp\Biggl[-\frac{1}{2}(\hat \phi_{A},\hat \phi_{B}) \left(\begin{array}{c}
 \hat  \phi_{A}\\
  \hat \phi_{B}
\end{array}\right)-\frac{1}{2}(\hat \phi'_{A},\hat \phi'_{B}) \left(\begin{array}{c}
 \hat  \phi'_{A}\\
  \hat \phi'_{B}
\end{array}\right)\no
&&\ \ \ \ \ \ \ -\frac{1}{4}\Big(\hat \phi_A-\hat \phi'_A,\hat \phi_B-\hat \phi'_B\Big)
\tilde N' \left(\begin{array}{c}
  \hat \phi_{A}-\hat \phi'_A \\
 \hat  \phi_{B}-\hat \phi'_B
\end{array}\right)
\Biggr], \nonumber\\\ea
where
\be
\tilde N'= M^{-1/2}_{D} V \tilde N V^{T} M_{D}^{-1/2}.
\ee
To diagonalize $\tilde N'$  we perform another transformation,
\be
\tilde \phi_{AB}= S \hat \phi_{AB},
\ee
where $S$ is another orthogonal matrix. Finally, up to a normalization factor we have
\ba
&& \rho^{\Gamma_{B}}_{AB}[(\tilde \phi_{A},\tilde \phi_{B}),(\tilde \phi'_{A},\tilde \phi'_{B})]\no
&&\propto \exp\Biggl[-\frac{1}{2}(\tilde \phi_{A}^2+\tilde \phi_B^2+(\tilde \phi'_{A})^2+(\tilde \phi'_{B})^2)\no
&&-\frac{1}{4}\Big(\tilde \phi_A-\tilde \phi'_A,\tilde \phi_B-\tilde \phi'_B\Big)
\hat N \left(\begin{array}{c}
  \tilde \phi_{A}-\tilde \phi'_A \\
 \tilde  \phi_{B}-\tilde \phi'_B
\end{array}\right)
\Biggr],  \label{www}
\ea
where
\be
\hat N=\begin{bmatrix}\mu_{1} & & \\ & \ddots & \\ & & \mu_{|A|+|B|}\end{bmatrix}.\ee
Here $\mu_i$ are the eigenvalues of the matrix $\tilde N'$, equivalently
the eigenvalues of the matrix  $M^{-1} \tilde N$. This is because
we can write $\tilde N'$ as $\tilde N'=(\s{M_D}V)\cdot  M^{-1}\ti{N}\cdot (\s{M_D}V)^{-1}$.

Once we numerically obtain these eigenvalues $\mu_i$ we can calculate the logarithmic negativity
in a similar way to  the entanglement entropy in
\cite{BKLS,Sha}. As a toy model, consider a scalar $\phi$ in
quantum mechanics with the density matrix
\be
\rho[\phi,\phi']\propto e^{-\frac{1}{2}(\phi^2+\phi'^2)-\frac{\mu}{4}(\phi-\phi')^2}. \label{qqq}
\ee
We can diagonalize (\ref{qqq}) and find the eigenvalues
\be
(1-\lambda)\lambda^m,\ \ \ (m=0,1,2,\ddd,\infty),
\ee
where $\lambda$ is defined by
\be
\lambda=\frac{2+\mu-2\s{1+\mu}}{\mu}.
\ee
Thus we obtain
\be
\log |\rho|= \log\left[(1-\lambda) \sum_{m=0}^\infty|\lambda|^m\right]=\log\frac{1-\lambda}{1-|\lambda|},
\ee
which is non-negative and is positive when $\mu$ is negative.

Now notice that our $\rho^{\Gamma_{B}}_{AB}$ given by (\ref{www}) can be regarded as $|A|+|B|$ copies of this kind of quantum mechanics. Thus, finally, we can evaluate the logarithmic negativity as follows:
\be
\mathcal{E}_{N}(\rho_{AB})=\sum_{i=1}^{|A|+|B|}\log\left[\frac{1-\lambda_i}{1-|\lambda_i|}\right],
\ee
where
\be
\lambda_i=\frac{2+\mu_i-2 \sqrt{1+\mu_i}}{\mu_i}.
\ee\\

\section{Appendix B: Computing EoP in Free Scalar Field Theory}\label{app:FS}

\subsection{EoP under $Z_2$ symmetry}

We first show that the entanglement of purification $E_P$ is invariant under a $Z_2$ symmetry transformation. In terms of the $w\times w$ matrix $K$ determining the coupling between the $AB$ and $\tilde{A}\tilde{B}$ systems, this symmetry is expressed as:
\begin{equation}
K \to K^R = S_{2w} K S_{2w} =
\begin{pmatrix}
S_{w} K_{B, \tilde{B}} S_{w} & S_{w} K_{B, \tilde{A}} S_{w} \\
S_{w} K_{A, \tilde{B}} S_{w} & S_{w} K_{A, \tilde{A}} S_{w}
\end{pmatrix}\ , 
\end{equation}
where $S_d$ is the $d \times d$ reversion matrix $(S_d)_{j,k} = \delta_{j,d+1-k}$ with $S=S^\mathrm{T}=S^{-1}$. For the $w \times w$ matrix $L$ determining the coupling within $\tilde{A}\tilde{B}$, we use \eqref{EQ_LMATRIX} to find
\begin{align}
L^{-1} &\to (S_{2w} K^{-1} S_{2w} Q) R^{-1} (S_{2w} K^{-1} S_{2w} Q)^\mathrm{T} \nonumber\\
&= S_{2w} L^{-1} S_{2w}\ ,
\end{align}
where we used $S_{2w} Q R^{-1} Q^\mathrm{T} S_{2w} = Q R^{-1} Q^\mathrm{T}$, as the initial system $A$ and $B$ is symmetric under the $Z_2$ symmetry. For the same reason, $J \to S_{2w} J S_{2w} $. As a result, the matrix $\Lambda$ from whose eigenvalue spectrum we compute $E_P$ becomes
\begin{align}
\Lambda &= -V^{-1}_{A\tilde{A},B\tilde{B}}\, V^{\phantom{-1}}_{B\tilde{B},A\tilde{A}} \nonumber\\
&\to -S_{2w} \, V^{-1}_{B\tilde{B},A\tilde{A}}\, V^{\phantom{-1}}_{A\tilde{A},B\tilde{B}}\, S_{2w} \ .
\end{align}
This is merely a similarity transform (since $S=S^{-1}$) and a transpose, which do not affect the eigenvalues. Hence, for any purification in $AB\tilde{A}\tilde{B}$ that is not itself $Z_2$-invariant, there exists another purification with identical EoP that is produced by acting with the $Z_2$ symmetry.

\begin{figure}[tb]
\centering
\includegraphics[width=3.6cm]{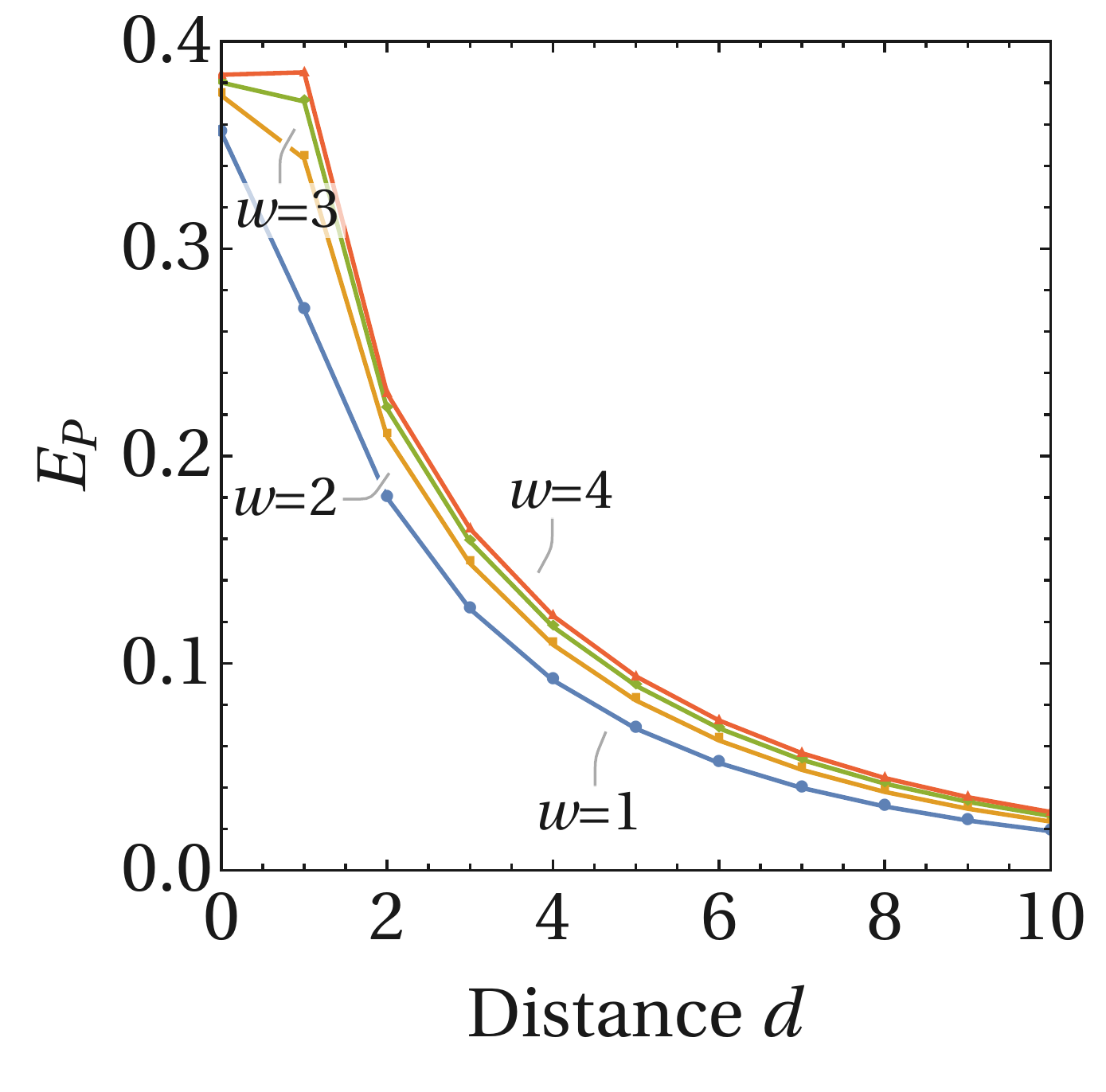}
\hspace{0.3cm}
\includegraphics[width=3.6cm]{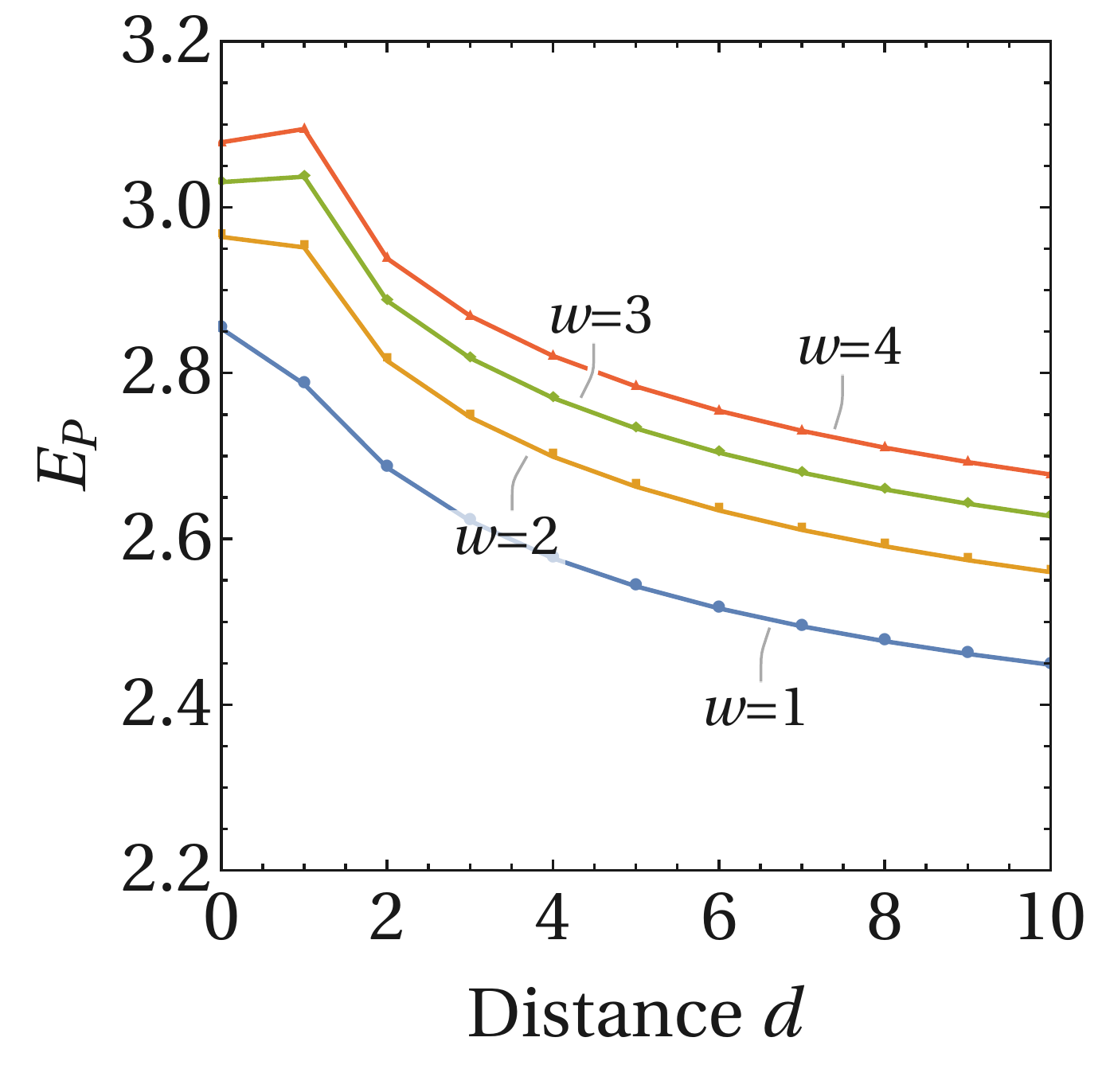}
\caption{Entanglement of purification $E_P$ for $w=|A|=|B|=1,2,3,4$ (blue, yellow, green, red) for two different 
masses $m=10^{-1}$ (left) and $m=10^{-4}$ (right), assuming a $Z_2$ coupling symmetry to the auxiliary system.
}
\label{FIG_EOP_Z2}
\end{figure}

A study of $Z_2$-invariant purifications for $w=1,2$ at $m \ll 1$ was already considered in \cite{Bhattacharyya:2018sbw}. Extending to $w=3,4$ yields the data shown in Fig.\ \ref{FIG_EOP_Z2}. We observe a peculiar peak of $E_P$ at $d=1$ appearing at larger $w$ and a power-law decay at $d>1$. As discussed in the main text, the peak disappears after relaxing the $Z_2$ constraint.

\subsection{Conformal limit}
\begin{figure}[tb]
\centering
\includegraphics[width=3.6cm]{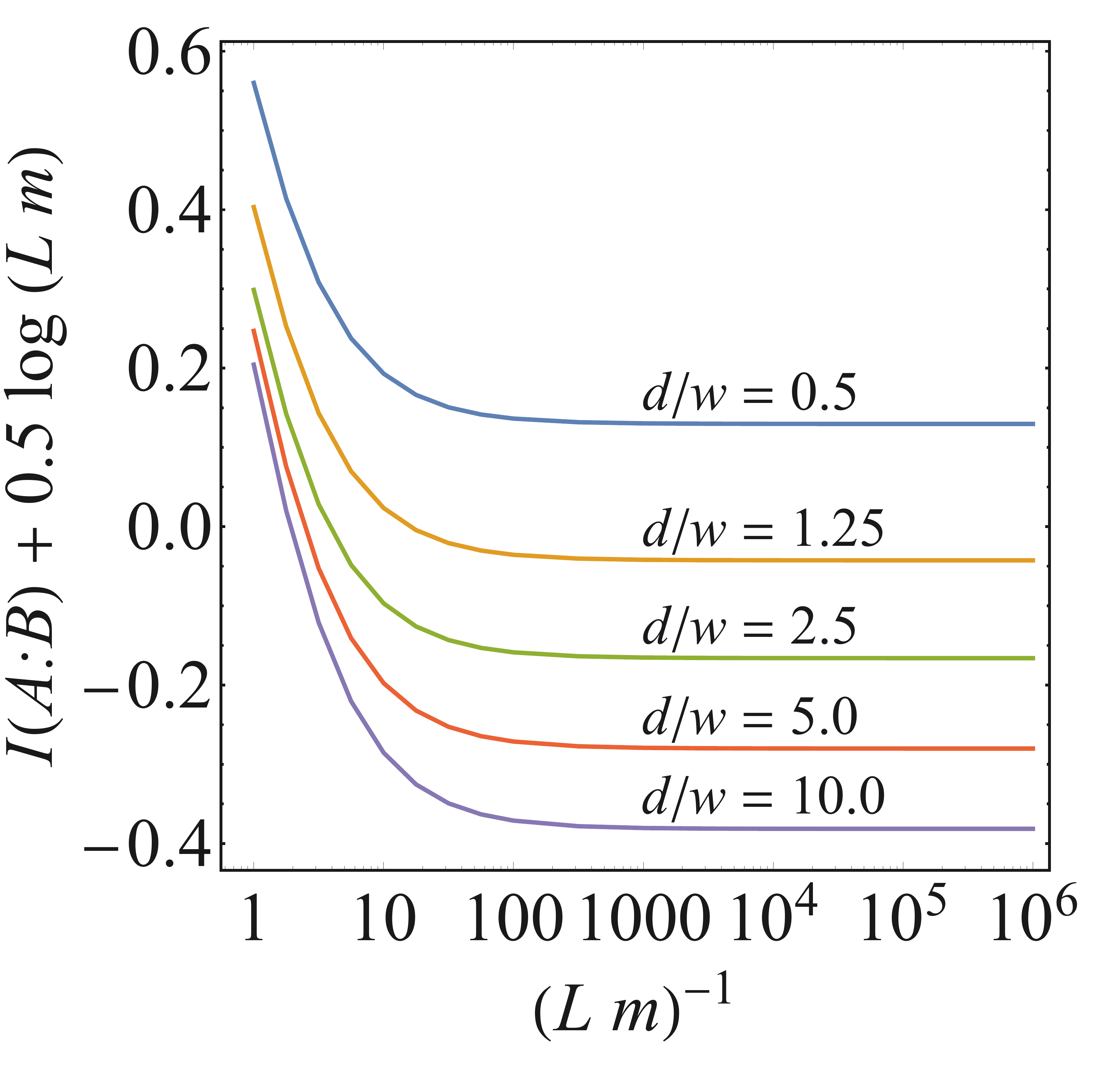}
\hspace{0.3cm}
\includegraphics[width=3.6cm]{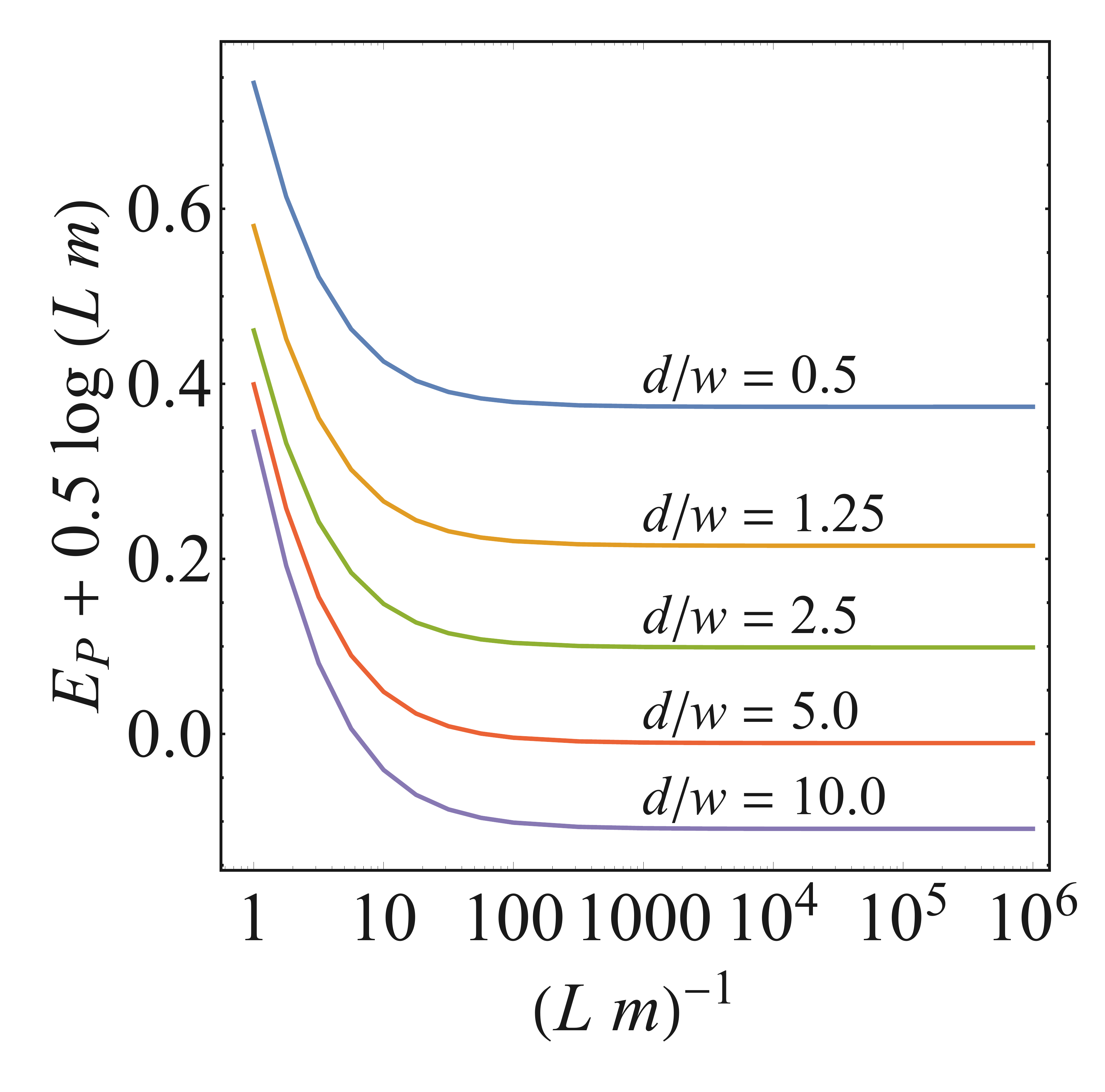}
\caption{\label{FIG_LM_SCALING} 
Scaling of mutual information $I(A:B)$ (left) and $E_P$ (right) with $L m$ at $w=4$. We add $\frac{1}{2} \log(L m)$ to each quantity and observe convergence to a constant at $L m \ll 1$, showing a $\propto - \frac{1}{2} \log(L m)$ dependence.}
\end{figure}

Our lattice formulation of both MI and EoP in scalar field theory relies on the following parameters: The lattice scaling $a$, the number of sites $N$, the mass $m$, the block widths $w$ (assuming MI/EoP between blocks $A$ and $B$ of equal width), and the distance $d$ between the blocks. The last two parameters are given in numbers of lattice sites. 
From the definition of the bosonic generating matrix $W$ (Eq.\ (5) in the main text) that determines the ground state wave function, we see that the system depends on the product $m a$ but not on $m$ or $a$ separately. 
In the conformal limit of an infinitely large and infinitely fine-grained system, we expect all observables to be invariant under two transformations: (a) The fine-graining transformation $N \to 2N, m a \to m a/2, d \to 2d, w \to 2w$ that replaces each lattice sites by two new ones. (b) The scaling transformation $d \to 2d, w \to 2w$ that rescales the subsystems $A$ and $B$.
Combining both conditions, an invariant quantity can depend only on the terms $d/w$ and $N m a = L m$, using the system length $L=N a$.
To determine the dependence of MI and EoP on these parameters, we perform numerical computations in the limits $L m \ll 1, w \gg 1,$ and $N \to \infty$. 
First, we consider the dependence on $L m$. As shown in Fig.\ \ref{FIG_LM_SCALING}, both MI and EoP show a clear $\propto - \frac{1}{2}\log(L m)$ behavior for $L m \ll 1$. As changing $d/w$ evenly shifts both values, we can already write
\begin{align}
\label{MI_EOP_FUNC_FORM}
I(A:B) &= - \frac{1}{2} \log(L m) + f(d/w)\ , \\
E_P &= - \frac{1}{2} \log(L m) + g(d/w)\ ,
\end{align}
in the conformal limit, with functions $f$ and $g$ to be determined next. 

To probe $d/w$ at both small and large values, we need to consider data series for different block widths $w$: As we need to make $N$ sufficiently large as $d$ increases to capture the $N \to \infty$ limit, we can compute the large $d/w$ range most easily at small $w$. While we expect finite-size effects at small $w$, we find that $w \leq 2$ already gives accurate results at $d/w \gg 1$. 
For the $d/w < 1$ range, we need to consider larger values of $w$, as we are constrained by $d \geq 2$ if we want to avoid the lattice effects explained in the main text. To reduce the computational cost of computing data at large $d$, we produce data sets for different $w$ and combine them. The greater computational cost of computing EoP at large $w$ means that we can study its small $d/w$ limit less effectively than for MI.

\begin{figure}[tb]
\centering
\includegraphics[width=6.2cm]{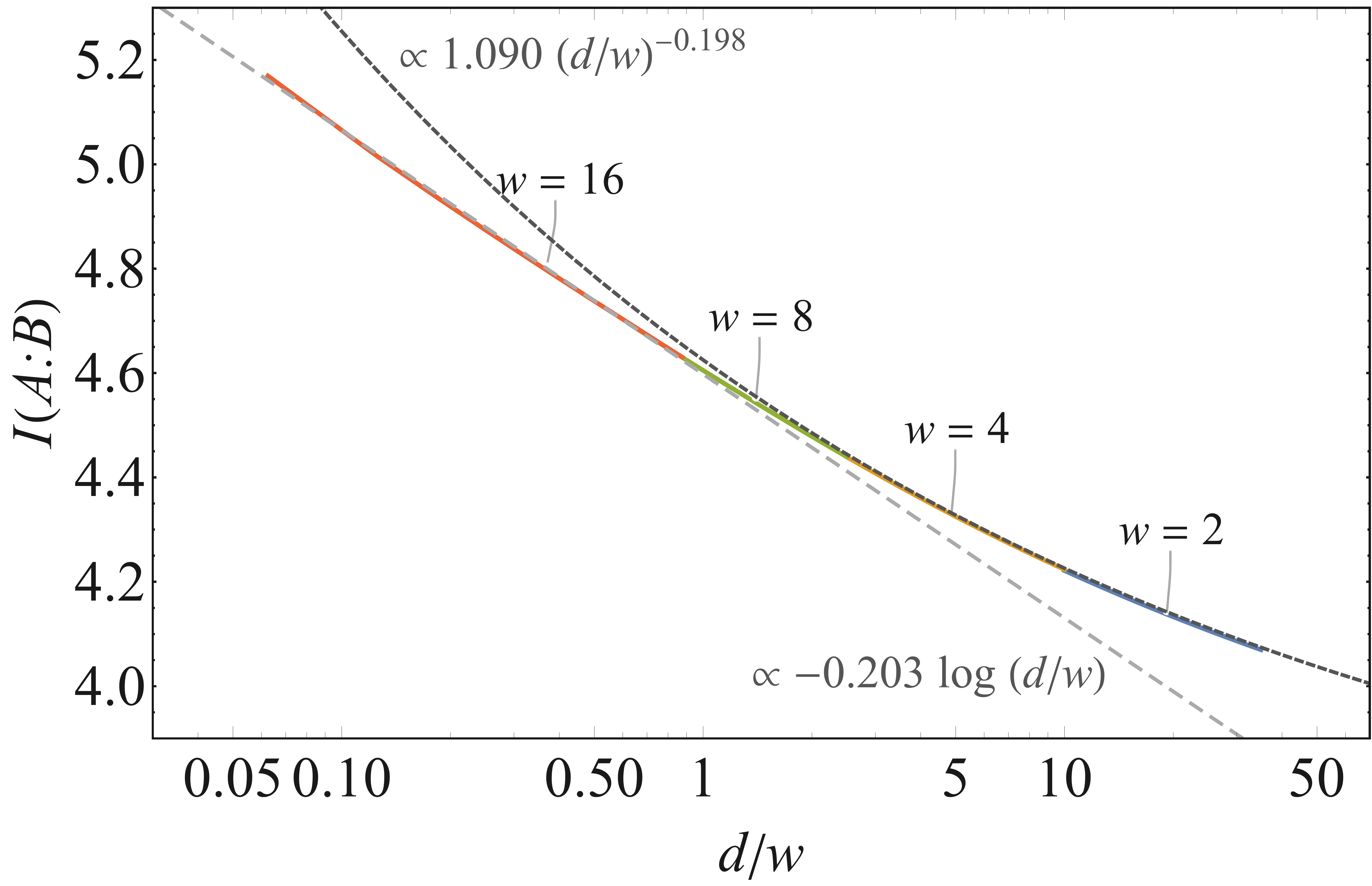}
\includegraphics[width=6.2cm]{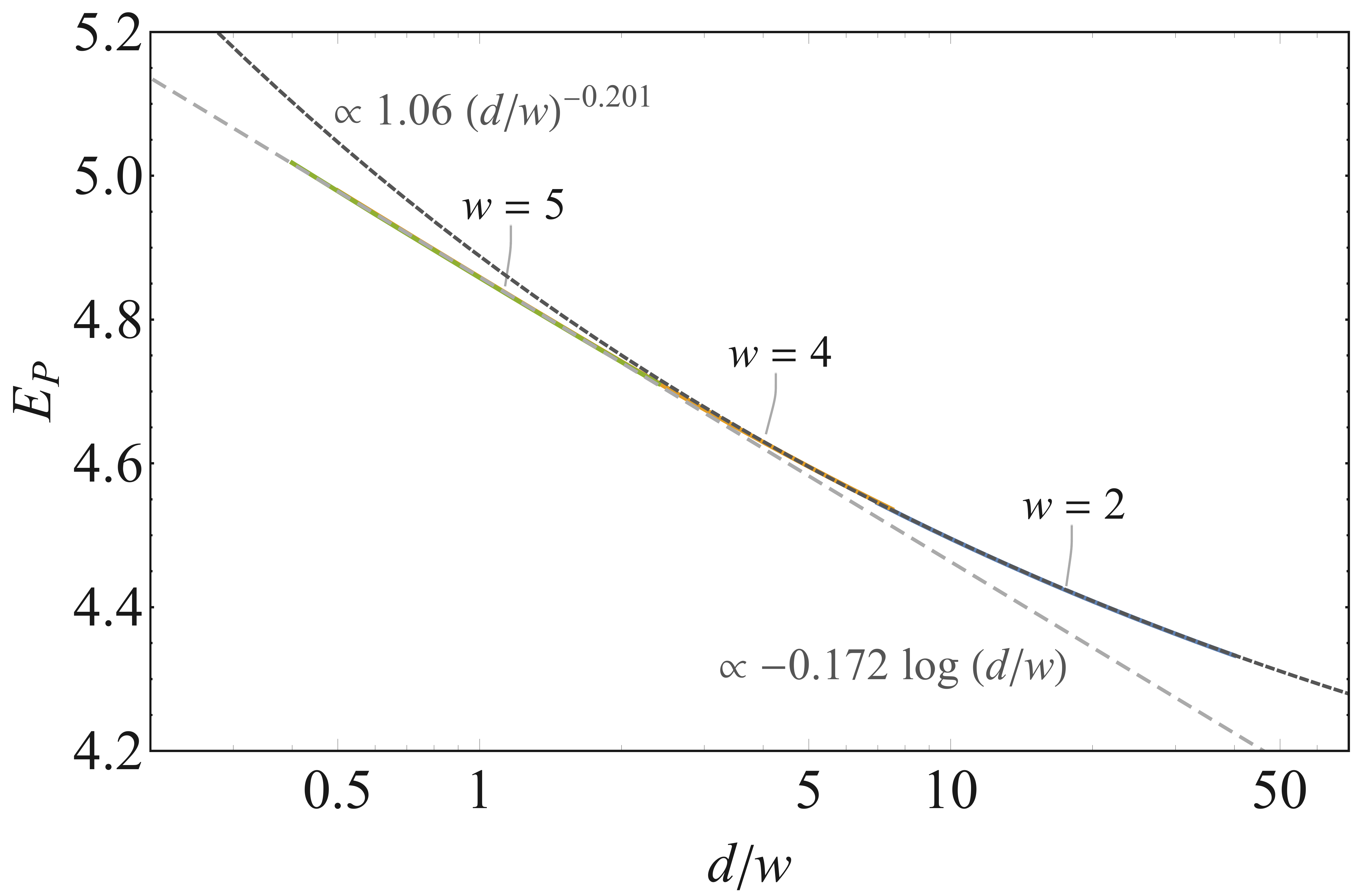}
\caption{\label{FIG_DW_SCALING} 
Mutual information $I(A:B)$ (top) and entanglement of purification $E_P$ (bottom) over a large range of $d/w$, composed of several data sets at $w=2,4,8,16$ for MI and $w=2,4,5$ for EoP, with $L m = 10^{-4}$. Logarithmic and power fit functions drawn as dashed and dotted curves, respectively.}
\end{figure}

The numerical results are shown in Fig.\ \ref{FIG_DW_SCALING}. We find a remarkably similar $d/w$ dependence of MI and EoP: Both scale logarithmically at $d/w \lesssim 1$ and turn into a power law at $d/w \gg 1$. We find the following fit functions $f$ and $g$ (following \eqref{MI_EOP_FUNC_FORM}) in both limits: 
\begin{align}
f(d/w) &\approx 
\begin{cases}
-0.007 - 0.203 \log\frac{d}{w} & \text{for }\frac{d}{w} \lesssim 1 \\
-1.07 + 1.09 \left( \frac{d}{w} \right)^{-0.198} & \text{for }\frac{d}{w} \gg 1
\end{cases}\ , \\
g(d/w) &\approx 
\begin{cases}
0.254 - 0.172 \log\frac{d}{w} & \text{for }\frac{d}{w} \lesssim 1 \\
-0.777 + 1.06 \left( \frac{d}{w} \right)^{-0.201} & \text{for }\frac{d}{w} \gg 1
\end{cases}\ .
\end{align}
The numerical results between MI and EoP only differ significantly in two regards: For $d/w \lesssim 1$ the logarithmic decay is slightly faster for MI than EoP, and in both ranges of $d/w$ both quantities differ by a constant $C \approx 0.25$.
 
\section{Appendix C: Computing EoP in Spin Systems}\label{app:Spin}

\subsection{Details of Numerical Calculation of EoP}

Let us briefly review the EoP in a finite-dimensional system. Given a bipartite state $\rho_{AB}$, any purification of $\rho_{AB}$ can be created from an arbitrary
initial purification $\ket{\psi_{0}}$ by acting with local unitary operations on the auxiliary system:
\begin{equation}
\ket{\psi(U_{\ti{A}\ti{B}})}_{AB\ti{A}\ti{B}}=I_{AB}\otimes U_{\ti{A}\ti{B}}\ket{\psi_{{\rm 0}}}_{AB\ti{A}\ti{B}}.
\end{equation}
As an initial purification, one may use the standard purification \cite{EP}
\begin{equation}
\ket{\psi_{{\rm std}}}_{AB\ti{A}\ti{B}}=\sum_{i}\sqrt{p_{i}}\ket{i}_{AB}\ket{i}_{\ti{A}}\ket{0}_{\ti{B}},
\end{equation}
or the TFD purification (\ref{TFD}). Note that we can regard $\ket{\psi_0}$ as a vector in higher-dimensional purification space, especially for the maximal purification $D_{\ti{A}\ti{B}}=({\rm rank} \rho_{AB})^2$, without loss of generality. Therefore, the minimization of EoP over all possible purifications can be equivalently expressed in terms of unitary operators on auxiliary systems:
\ba
&& E_{P}(\rho_{AB})  =\min_{U_{\ti{A}\ti{B}},\ \ti{A}}S(\rho_{A\ti{A}}), \no
\rho_{A\ti{A}}&&={\rm Tr}_{B\ti{B}}[I_{AB}\otimes U_{\ti{A}\ti{B}}\ket{\psi_{0}}\bra{\psi_{0}}I_{AB}\otimes U_{\ti{A}\ti{B}}^{\dagger}],
\ea
where the minimization is also taken over all possible divisions of
the ancilla Hilbert space into $\mathcal{H}_{\ti{A}}$ and $\mathcal{H}_{\ti{B}}$,
imposing $D_{\ti{A}}D_{\ti{B}}\leq{\rm (rank}\rho_{AB})^{2}$. Note that the optimal purification has a trivial redundancy since $S_{A\ti{A}}$
is invariant under any local unitary $U_{\ti{A}\ti{B}}=U_{\ti{A}}\otimes U_{\ti{B}}$. 
In other words, $S_{\ti{A}}\neq S_{\ti{B}}$ indicates a non-trivial degeneracy of the optimal purification.

\subsection{Werner state}

An interesting type of quantum state is
the Werner state on 2 qubits system
\begin{align}
\rho_{AB}(p) & =\frac{p}{3}P_{{\rm sym}}+(1-p)P_{{\rm asym}}\\
 & =\frac{p}{3}I+(1-\frac{4p}{3})\ket{\rm Bell}\bra{\rm Bell},
\end{align}
where $p\in[0,1]$ is a parameter of states, $P_{{\rm sym}}$ and
$P_{{\rm asym}}$ are the projections onto the (anti-)symmetric subspace in $H_{AB}$
\begin{equation}
P_{{\rm sym}}=\begin{pmatrix}1\\
 & \frac{1}{2} & \frac{1}{2}\\
 & \frac{1}{2} & \frac{1}{2}\\
 &  &  & 1
\end{pmatrix},\ P_{{\rm asym}}=\begin{pmatrix}0\\
 & \frac{1}{2} & -\frac{1}{2}\\
 & -\frac{1}{2} & \frac{1}{2}\\
 &  &  & 0
\end{pmatrix},
\end{equation}
in $ \{ \ket{00},\ket{01},\ket{10},\ket{11}\}$ basis, $I$ is the $4\times 4$ identity matrix, and $\ket{\rm Bell}:=\frac{1}{\sqrt{2}}(\ket{01}-\ket{10})$. 
The Werner state is also related to an isotropic state,
\begin{equation}
\rho_{AB}(p)=\frac{I}{4}+\frac{1}{4}(\frac{4p}{3}-1)\sum_{i=x,y,z}
\sigma_{A}^{i}\otimes\sigma_{B}^{i},\label{eq:WernerPauli}
\end{equation}
which can appear as the ground state of the anti-ferromagnetic Heisenberg model.

The EoP of the Werner state has already been calculated numerically in \cite{EP, Wint} (using a slightly different  
definition $\ket{\rm Bell}:=\frac{1}{\sqrt{2}}(\ket{00}+\ket{11})$, which does not change the correlation). 
Here, we computed the EoP for Werner state again and found more fine-grained phase structures related to the $Z_2$ symmetry breaking.


\begin{figure}
\centering
\includegraphics[height=3.6cm]{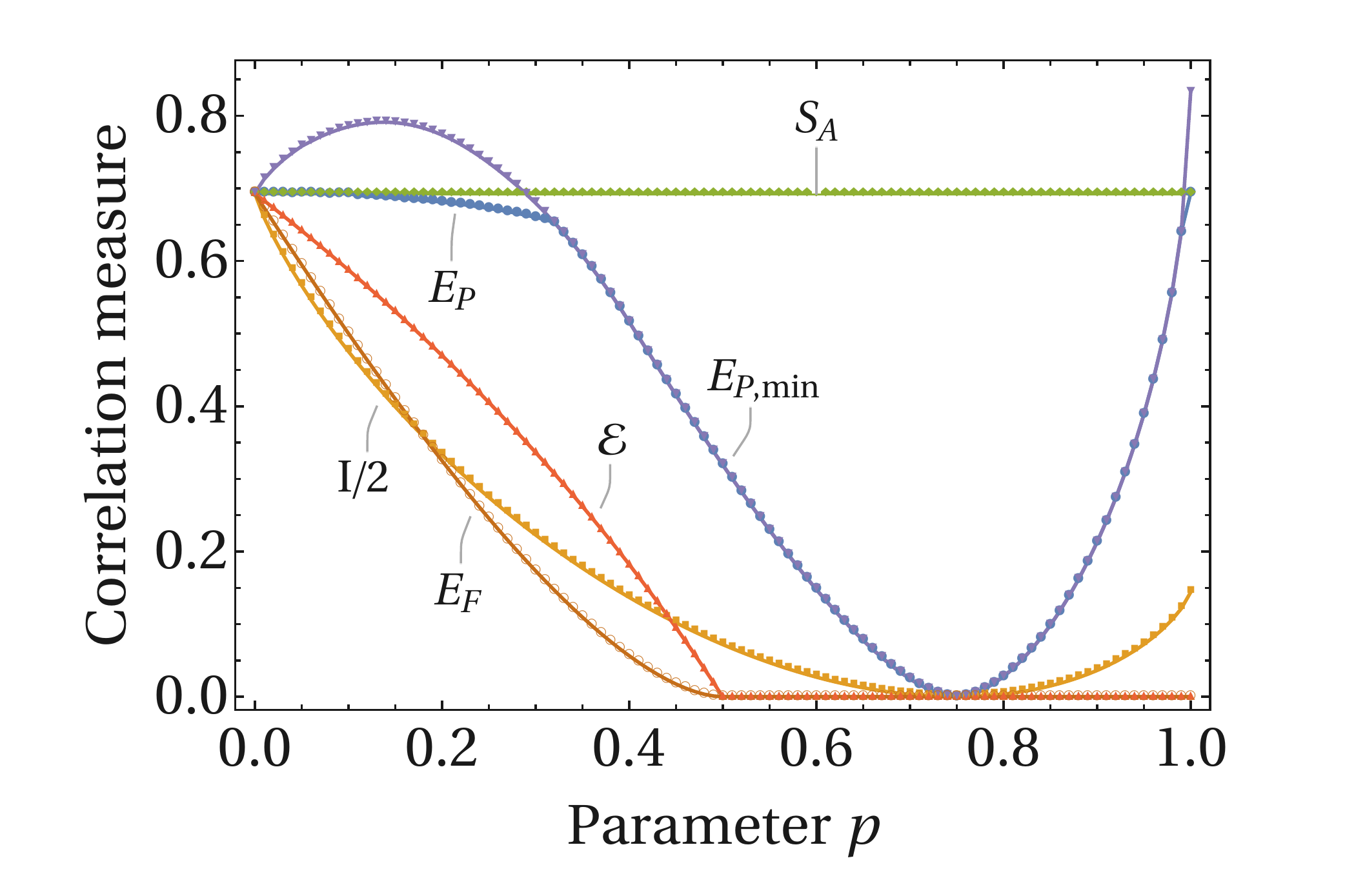}
\caption{\label{fig:EoPWerner}The EoP $E_P$ for Werner
states. We also show the entanglement of formation $E_F$ (obtained from the formula \cite{EoFqubits}), log negativity $\varepsilon$, $I(A:B)/2$, as well as the EoP for minimal purifications $E_{P,\text{min}}$. }
\end{figure}

\begin{figure}[tb]
\centering
\includegraphics[width=3.5cm]{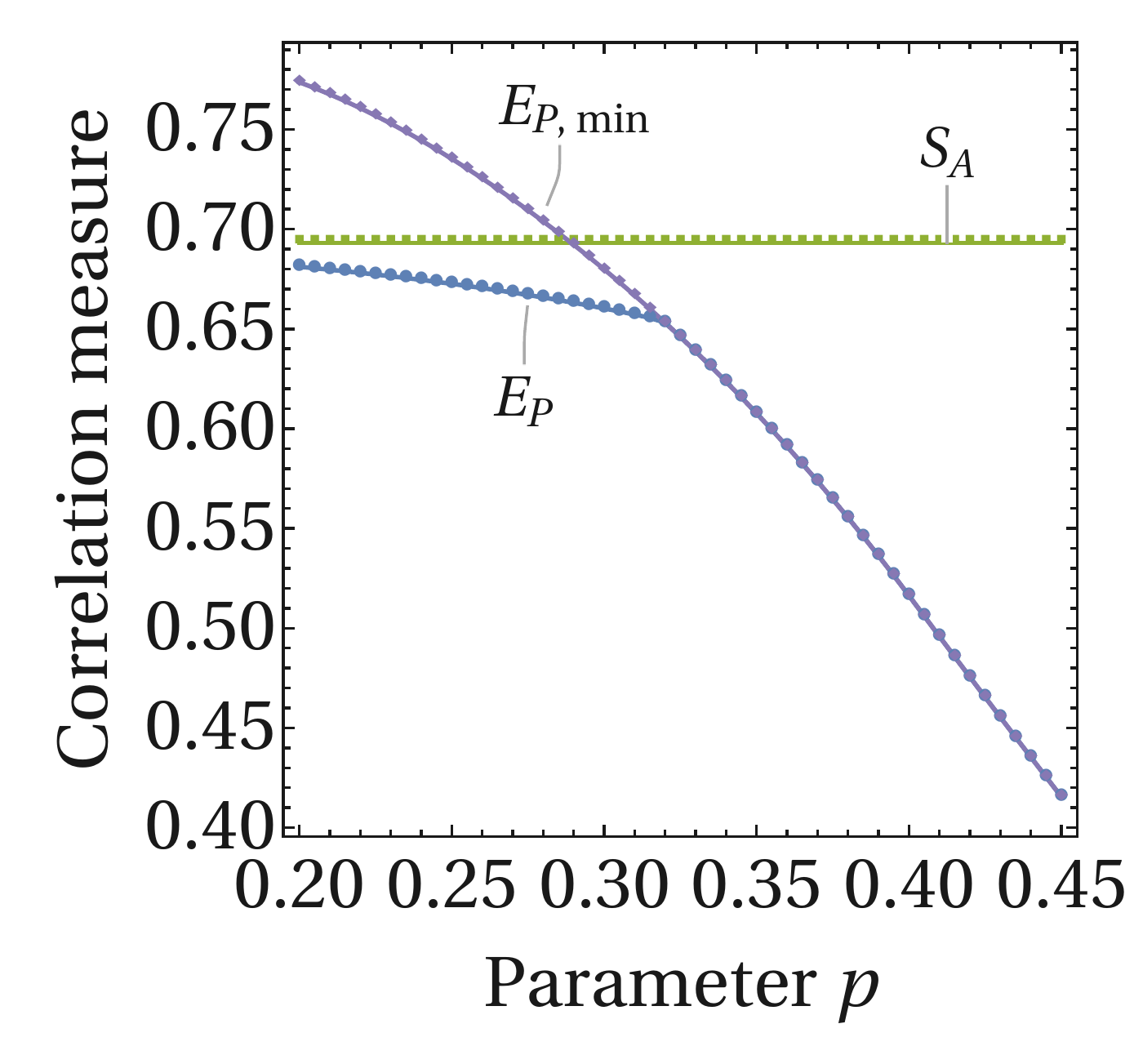}
\hspace{0.3cm}
\includegraphics[width=3.5cm]{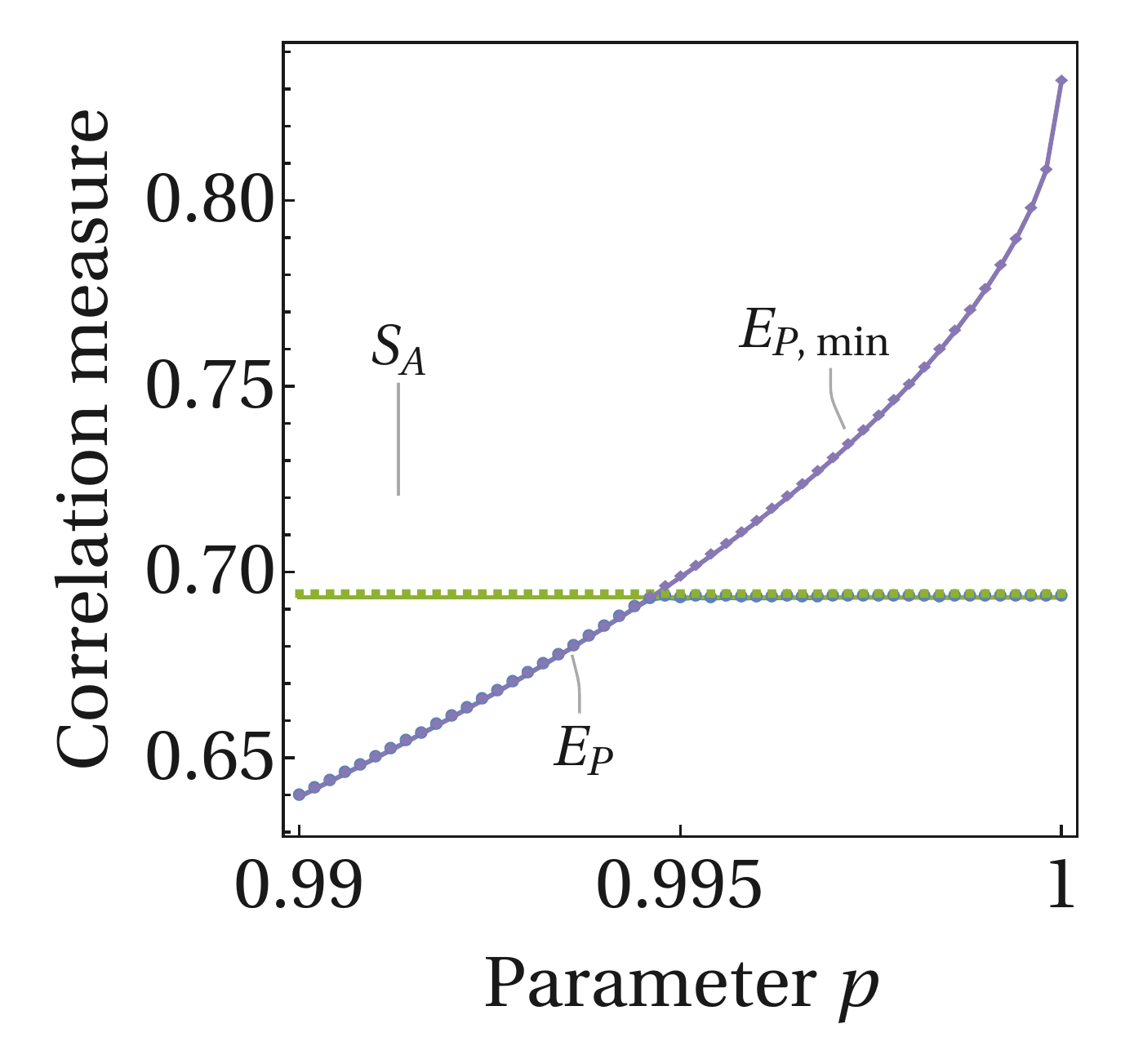}
\caption{\label{figEoPWernerPTs}
The EoP for Werner states for (left) $0.2\protect\leq p\protect\leq0.45$ and (right) $0.99\protect\leq p\protect\leq1$. There is a small gap $S_A-E_P>0$ even in the phase (d) (except $p=1$) though it is hard to observe numerically.}
\end{figure}

The result is shown in Fig.\ \ref{fig:EoPWerner}. There are 4 different
regimes classified by a configuration of optimal purifications: (a) A non-minimal
purification phase $0\leq p\lesssim0.319$ with $D_{\ti{A}}=2$, $D_{\ti{B}}=3$ (or $D_{\ti{A}}=D_{\ti{B}}=3$),
(b) A minimal purification phase $0.319\lesssim p\lesssim0.401$ with
$D_{\ti{A}}=D_{\ti{B}}=2$, (c) TFD purification phase $0.401\lesssim p\lesssim0.995$
where the optimal purification is given by the TFD purification, and (d) A non-minimal purification
phase $0.995\lesssim p\lesssim1$ with $D_{\ti{A}} =2$, $D_{\ti{B}}=4$. These phase transitions are depicted in Fig. \ref{figEoPWernerPTs}.

In the phase (a), we found that there are two choices (non-equivalent) for the optimal purifications: $D_{\ti{A}}=2, D_{\ti{B}}=3$ and $D_{\ti{A}}=D_{\ti{B}}=3$ which produces the same results for EoP up to certain the numerical accuracy. Indeed, they give the same value for $S_{A\ti{A}}$ (after minimization) up to 10 digits around the transition point. For the results shown in the figure, we have used  $D_{\ti{A}}=2, D_{\ti{B}}=3$. In the phase (d), it was observed that the EoP is strictly smaller than $S_A$ (except $p=1$), which was missed in the previous works. 

\begin{figure}[tb]
\centering
\includegraphics[width=3.5cm]{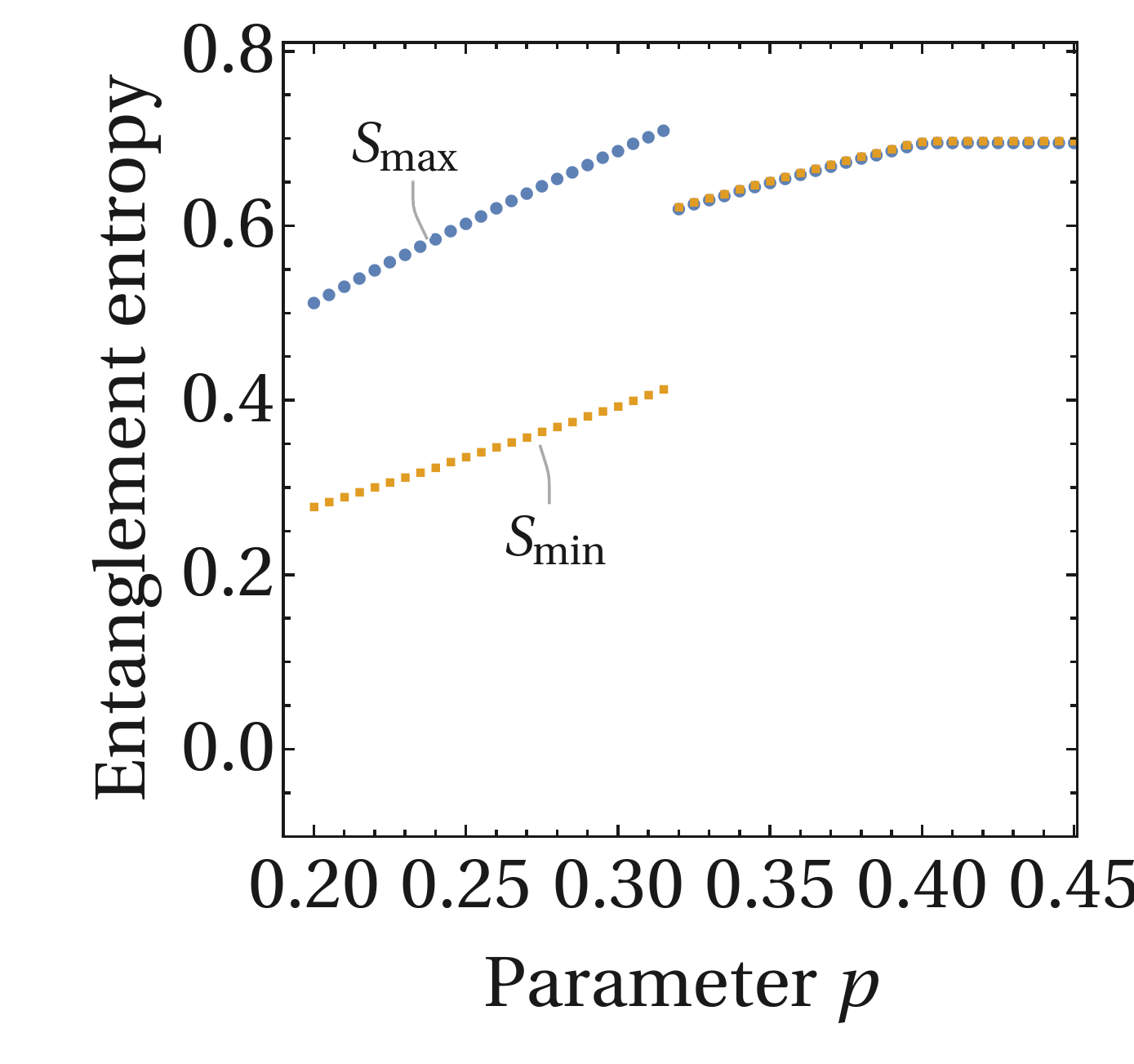}
\hspace{0.3cm}
\includegraphics[width=3.5cm]{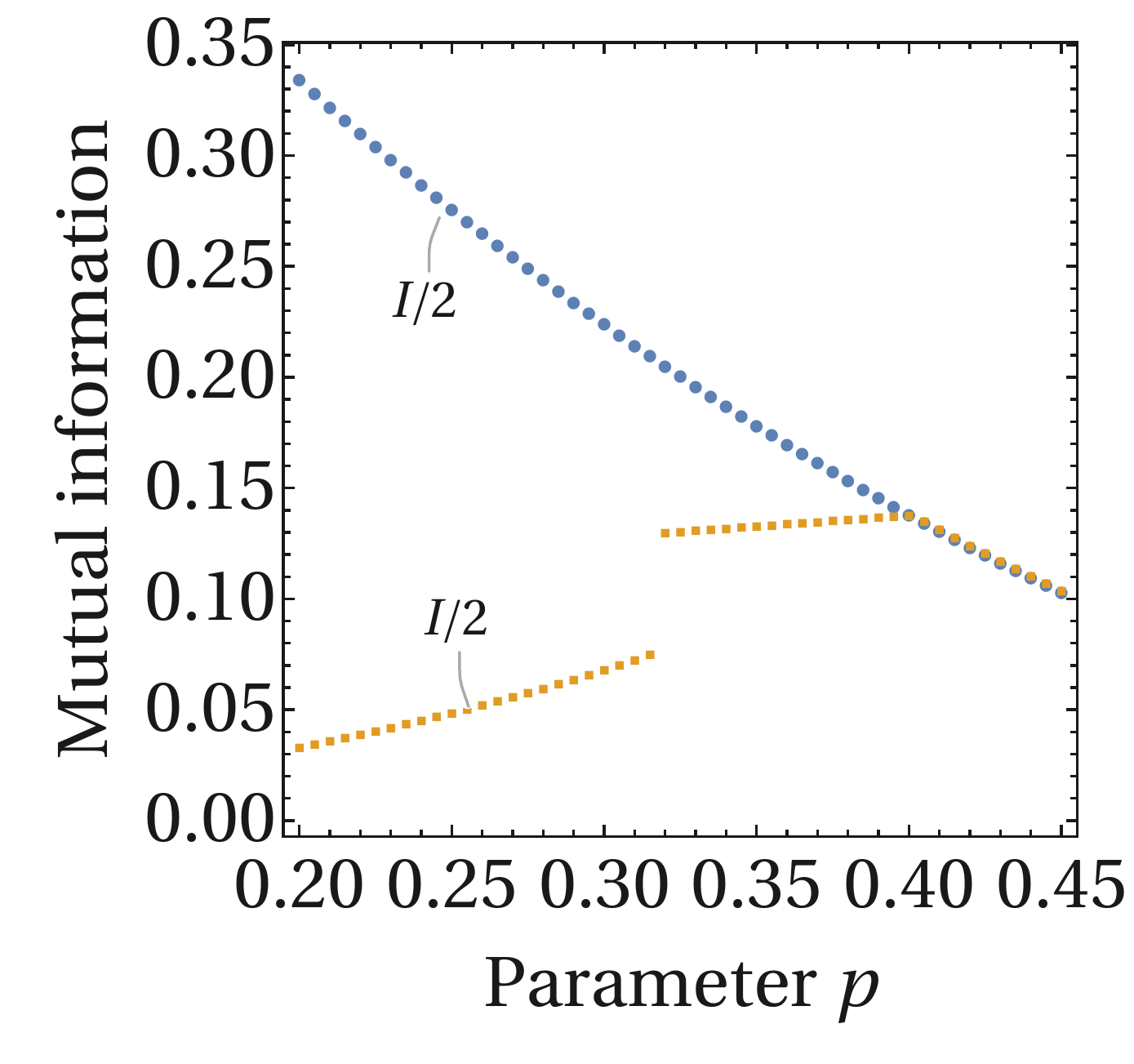}
\caption{\label{fig:EoPWernerPTSApSBpIApBp}
The plots of $S_{\ti{A}},\ S_{\ti{B}}$ (left), and $I(\ti{A}:\ti{B})$ (right) of the optimal purification around $0.2\protect\leq p\protect\leq0.45$.
There are two phase transition points at $p_{1}\simeq0.319$ and $p_{2}\simeq0.401$,
which separate the phases (a) from (b), and (b) from (c), respectively. As we are using the $D_{\ti{A}}<D_{\ti{B}}$ setup, $S_{\text{min}}=S_{\ti{A}}, S_{\text{max}}=S_{\ti{B}}$.}
\end{figure}

\begin{figure}[tb]
\centering
\includegraphics[width=3.5cm]{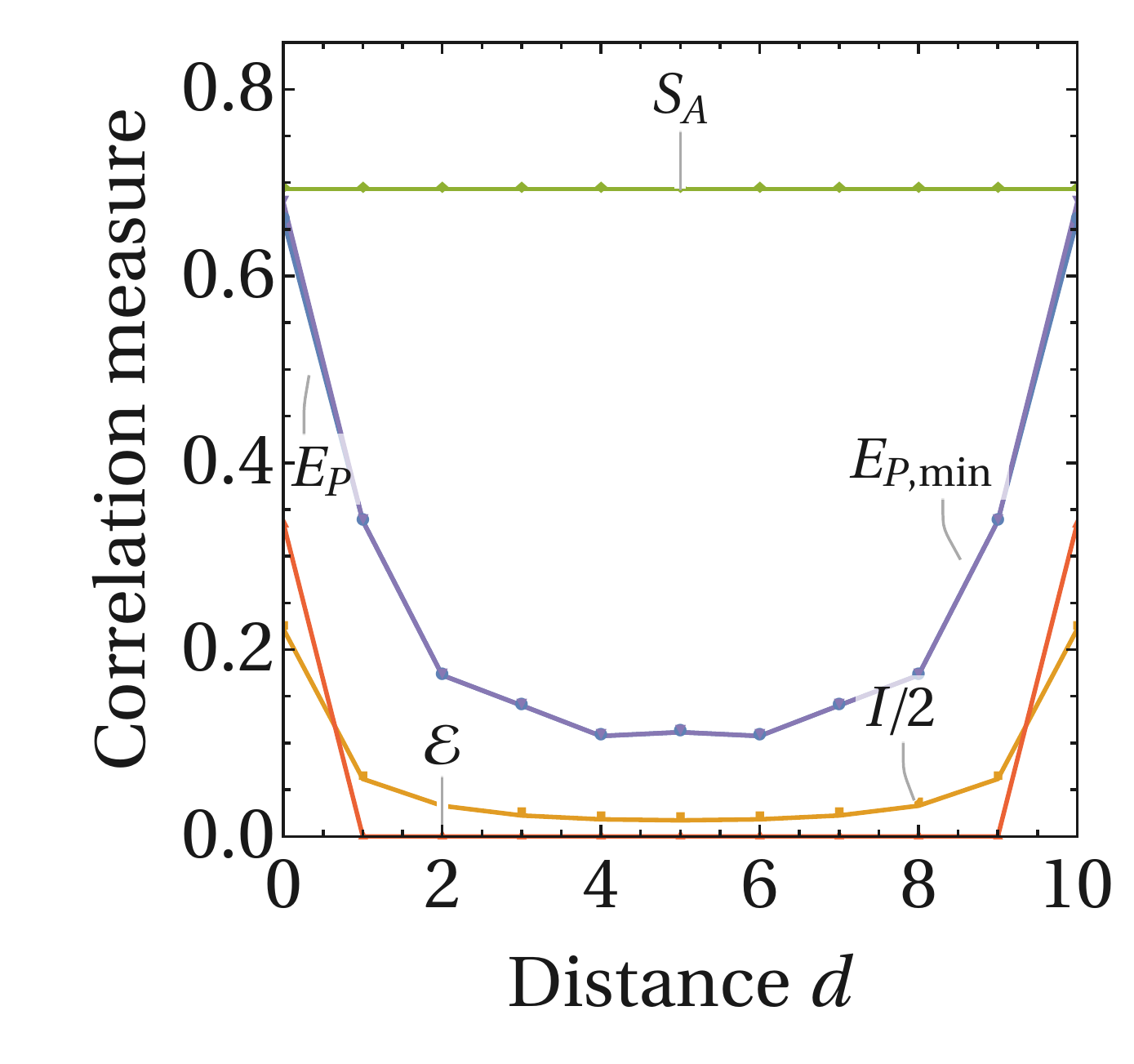}
\hspace{0.3cm}
\includegraphics[width=3.5cm]{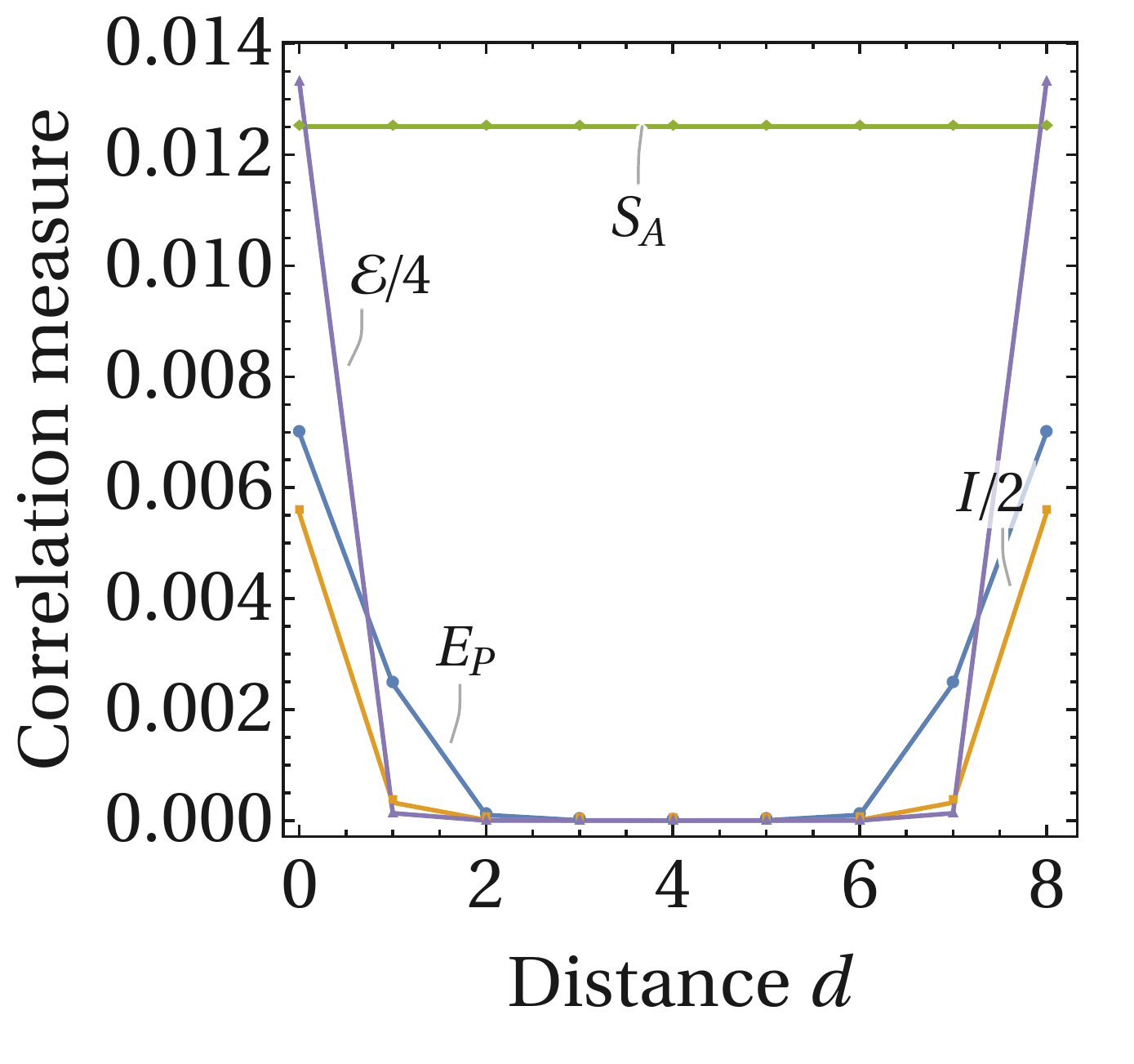}
\caption{\label{fig:EoPHeisenbergAndChaos} The entanglement of purification for the anti-ferromagnetic isotropic Heisenberg model (left) and for a chaotic spin chain (right).}
\end{figure}

Now let us focus on the optimal purifications around the transition points $p_{1}\simeq0.319$ and $p_{2}\simeq0.401$. $S_{\ti{A}},\ S_{\ti{B}}$ and $I(\ti{A}:\ti{B})$ around this region are shown in Fig.\ \ref{fig:EoPWernerPTSApSBpIApBp}.
It is clear that the optimal purification breaks the $Z_2$ reflection symmetry which exchanges $\ti{A}$ and $\ti{B}$ in the phase (a). It is analogous to what we found in the free scalar field theory and in the critical Ising spin chain. Note that $S_{\ti{A}}=S_{\ti{B}}$ holds for phase (b) and (c), contrary to the Ising model.  We also find that $Z_2$ reflection symmetry is similarly broken in the phase (d).

\subsection{Heisenberg model}

Let us consider the anti-ferromagnetic isotropic Heisenberg model
\begin{equation}
H_{{\rm Heisenberg}}=\sum_{\langle i,j \rangle}(\sigma_{i}^{x}\otimes\sigma_{j}^{x}+\sigma_{i}^{y}\otimes\sigma_{j}^{y}+\sigma_{i}^{z}\otimes\sigma_{j}^{z}).
\end{equation}
For even $N$, the reduced density matrix $\rho_{AB}$ of size $|A|=|B|=1$ constructed from  the ground state is equivalent to the Werner state \eqref{eq:WernerPauli} \cite{WernerHeisenberg}. We set the total number of spins to $N=12$. Interestingly, the resulting EoP exhibits a slight peak at the farthest distance $d=5$ while MI monotonically decreases (Fig.\ \ref{fig:EoPHeisenbergAndChaos}). This peak also shows that the EoP does not necessarily monotonically decrease along with other correlation measures.


\subsection{Chaotic spin chain}

Finally we consider a non-integrable model by adding a parallel magnetic field to the Ising model,
\begin{equation}
H_{{\rm Ising}}=-\sum_{\langle i,j \rangle}\sigma_{i}^{z}\otimes\sigma_{j}^{z}-h\sum_{i=1}^{N}\sigma_{i}^{x}-g \sum_{i=1}^{N}\sigma_{i}^{z}.
\end{equation}
We set the parameters $h=1.05$ and $g=-0.5$ following \cite{Chaos}. We use the same setup as the Heisenberg model and  find that the long range correlations are almost vanishing in the vacuum (Fig.\ \ref{fig:EoPHeisenbergAndChaos}).

\subsection{Non-monotonicity of spin chain EoP with $N=4$ }

The non-monotonicity of EoP for $N=4$ (Fig.\ \ref{fig:EoPIsing}) is common in {\it any} homogeneous spin chain. The key observation is that, when $N=4$, the symmetric and anti-symmetric projectors $P_{{\rm sym}}, P_{{\rm asym}}$ acting on $A$ and $B$ located on the diagonal position ($d=1$) are symmetries of the system, i.e.\ they commute with the Hamiltonian $[H, P_{{\rm sym}}]=[H, P_{{\rm asym}}]=0$. Indeed,  each term $\sum_{<ij>}\sigma_{i}^{l}\otimes\sigma_{j}^{l}, \sum_{i}\sigma_{i}^{l}$ ($l=x,y,z$) commutes with these projectors, and thence they are the symmetries of the system regardless of the values of coupling parameters. Since $P_{{\rm sym}}$ and $P_{{\rm asym}}$ are orthogonal, its unique vacuum (and any other non-degenerate excited state)  always  belongs to  either  symmetric or anti-symmetric subspace of $\mathcal{H}_{AB}$ (for example, in the vacuum of the anti-ferromagnetic isotropic Heisenberg model, $\rho_{AB}$ is $P_{{\rm sym}}/3$ itself). 

On the other hand, the EoP coincides with $S_A$ when $\rho_{AB}$ has support  either on the symmetric or on the antisymmetric subspace of $\mathcal{H}_{AB}$ \cite{Lock}. Thus we have $E_P=S_A$ at $d=1$, while $E_P<S_A$ at $d=0, 2$ in general, leading to a non-monotonic behavior of the EoP.

\end{document}